\documentclass[a4paper,UKenglish,cleveref, autoref]{lipics-v2019}

\usepackage{float}
\usepackage{hyperref}
\usepackage[utf8]{inputenc}
\usepackage{amsmath}
\usepackage{amssymb}
\usepackage{todonotes}
\usepackage{booktabs}
\usepackage{xspace}
\usepackage{placeins}

\DeclareMathOperator*{\flowto}{flow-to}
\DeclareMathOperator*{\flowfrom}{flow-from}

\DeclareMathOperator*{\cut}{cut}
\renewcommand{\epsilon}{\varepsilon}

\newcommand{\ie}{i.\,e.\xspace}
\newcommand{\eg}{e.\,g.\xspace}
\newcommand{\etal}{et al.\xspace}

\title{Evaluation of a Flow-Based Hypergraph Bipartitioning Algorithm} 

\titlerunning{Evaluation of a Flow-Based Hypergraph Bipartitioning Algorithm}

\author{Lars Gottesbüren}{Institute of Theoretical Informatics, Karlsruhe Institute of Technology, Germany}{lars.gottesbueren@kit.edu}{}{}

\author{Michael Hamann}{Institute of Theoretical Informatics, Karlsruhe Institute of Technology, Germany}{michael.hamann@kit.edu}{}{}

\author{Dorothea Wagner}{Institute of Theoretical Informatics, Karlsruhe Institute of Technology, Germany}{dorothea.wagner@kit.edu}{}{}

\authorrunning{L. Gottesbüren, M. Hamann, D.Wagner}

\Copyright{Lars Gottesbüren, Michael Hamann, Dorothea Wagner}

\ccsdesc[100]{Mathematics of computing~Graph algorithms}

\keywords{Hypergraph Partitioning, Maximum Flows, Algorithm Engineering}

\category{}

\supplement{Source code at \url{https://github.com/kit-algo/HyperFlowCutter}}

\funding{This work was partially supported by the DFG under grants WA654/19-2 and WA654/22-2. The authors acknowledge support by the state of Baden-W\"urttemberg through bwHPC.}

\acknowledgements{We thank Ben Strasser, Tim Zeitz and Lukas Barth for helpful discussions.}

\nolinenumbers 

\hideLIPIcs  

\EventEditors{John Q. Open and Joan R. Access}
\EventNoEds{2}
\EventLongTitle{42nd Conference on Very Important Topics (CVIT 2016)}
\EventShortTitle{CVIT 2016}
\EventAcronym{CVIT}
\EventYear{2016}
\EventDate{December 24--27, 2016}
\EventLocation{Little Whinging, United Kingdom}
\EventLogo{}
\SeriesVolume{42}
\ArticleNo{23}

\EventEditors{Michael A. Bender, Ola Svensson, and Grzegorz Herman}
\EventNoEds{3}
\EventLongTitle{27th Annual European Symposium on Algorithms (ESA 2019)}
\EventShortTitle{ESA 2019}
\EventAcronym{ESA}
\EventYear{2019}
\EventDate{September 9--11, 2019}
\EventLocation{Munich/Garching, Germany}
\EventLogo{}
\SeriesVolume{144}
\ArticleNo{49}

\begin{document}

\maketitle

\begin{abstract}
	In this paper, we propose HyperFlowCutter, an algorithm for balanced hypergraph bipartitioning.
	It is based on minimum $S$-$T$ hyperedge cuts and maximum flows.
	It computes a sequence of bipartitions that optimize cut size and balance in the Pareto sense, being able to trade one for the other.
	HyperFlowCutter builds on the FlowCutter algorithm for partitioning graphs.
	We propose additional features, such as handling disconnected hypergraphs, novel methods for obtaining starting $S,T$ pairs as well as an approach to refine a given partition with HyperFlowCutter.
	Our main contribution is ReBaHFC, a new algorithm which obtains an initial partition with the fast multilevel hypergraph partitioner PaToH and then improves it using HyperFlowCutter as a refinement algorithm.
	ReBaHFC is able to significantly improve the solution quality of PaToH at little additional running time.
	The solution quality is only marginally worse than that of the best-performing hypergraph partitioners KaHyPar and hMETIS, while being one order of magnitude faster.
	Thus ReBaHFC offers a new time-quality trade-off in the current spectrum of hypergraph partitioners.
	For the special case of perfectly balanced bipartitioning, only the much slower plain HyperFlowCutter yields slightly better solutions than ReBaHFC, while only PaToH is faster than ReBaHFC.
\end{abstract}

\section{Introduction}\label{sec:introduction}
Given a hypergraph $H=(V,E)$, a hyperedge cut $C \subset E$ is a set of hyperedges whose removal disconnects $H$.
The \emph{balanced hypergraph bipartioning problem} is to find a hyperedge cut of minimum cardinality whose removal separates $H$ into two blocks of roughly equal size -- up to $(1+\epsilon)\frac{|V|}{2}$.
Hypergraph partitioning has applications in VLSI design, database sharding, and high performance computing, in particular load balancing and reducing communication for highly parallel computations as well as accelerating sparse matrix vector multiplications.
This problem is NP-hard~\cite{l-caicl-90} and it is hard to approximate, even for graphs~\cite{bj-fnp-92}.
Therefore, practical algorithms use heuristics.
Most of them are based on the \emph{multilevel} framework~\cite{hl-amapg-95}.

In this paper we consider a different approach based on the max-flow min-cut duality.
The basic idea has already been used in the Flow-Balanced-Bipartition algorithm (FBB) of Yang and Wong~\cite{yw-enfbm-96}.
So far it has not been of further consideration due to being too slow compared to multilevel algorithms and too slow to solve current instances in feasible time.
More recently, FlowCutter~\cite{hs-gbpo-18} (FC) for graph bipartitioning has been introduced independently of FBB\@.
It is designed for computing very small node separators in road networks with a rather loose balance constraint; $\epsilon=0.33$ is recommended for the application of accelerating shortest path computations~\cite{dsw-cch-15}.
Based on similar ideas as FBB but equipped with more engineering, it computes both unbalanced and highly balanced bipartitions of high quality on the Walshaw graph partitioning benchmark~\cite{swc-tgpa-04}.

\subparagraph{Contribution.}
We present HyperFlowCutter, an algorithm which computes a series of hypergraph bipartitions with increasing balance, up to perfectly balanced solutions.
With HyperFlowCutter, we extend FlowCutter to hypergraphs and contribute additional features.
We provide a method to handle disconnected hypergraphs, which FlowCutter and FBB cannot handle.
Our main contribution is ReBaHFC, an algorithm to refine a given partition using HyperFlowCutter.
It is a natural extension of the max-flow based refinement of the k-way hypergraph partitioner KaHyPar~\cite{hss-nfbrm-18}.
We provide a thoroughly engineered implementation as well as an extensive experimental evaluation on the benchmark set of Heuer and Schlag~\cite{hs-icshp-17}, comparing HyperFlowCutter and ReBaHFC against the state-of-the-art hypergraph partitioning tools KaHyPar~\cite{hs-icshp-17,hss-nfbrm-18}, PaToH~\cite{patoh} and hMETIS~\cite{kaks-mhpav-99,kk-mkwhp-99}.

In our experiments we use the fast algorithm PaToH to obtain initial partitions for ReBaHFC.
When using the quality preset of PaToH, ReBaHFC computes solutions for $\epsilon = 0.03$, which are only slightly worse than those of the best-performing partitioners KaHyPar and hMETIS and significantly better than those of PaToH.
ReBaHFC is only marginally slower than PaToH and thus, like PaToH, it is one order of magnitude faster than KaHyPar and hMETIS, when using the quality preset, and two orders of magnitude faster, when using the default preset.
Furthermore, ReBaHFC with the PaToH default preset computes better solutions than PaToH with its quality preset.
Thus ReBaHFC offers new time-quality trade-offs.
For the special case of perfectly balanced bipartitioning, only the much slower plain HyperFlowCutter yields marginally better solutions than ReBaHFC, while only PaToH is faster than ReBaHFC.

\subparagraph{Outline.}
After discussing related work in Section~\ref{sec:related} and presenting notation and preliminaries in Section~\ref{sec:preliminaries}, we introduce the core algorithm of HyperFlowCutter for $S$-$T$ hyperedge cuts in Section~\ref{sec:core}.
Then we show how to handle disconnected hypergraphs in Section~\ref{sec:disconnectedhypergraphs}, propose our refinement algorithm ReBaHFC in Section~\ref{sec:refinement} and finally discuss the experimental evaluation in Section~\ref{sec:experiments}.

\section{Related Work}\label{sec:related}
For an overview of the field of hypergraph partitioning we refer to survey articles~\cite{bmsw-gpgcd-13,pm-hpc-07,ak-rdnps-95}.
The most common approach among hypergraph partitioning tools is the multilevel framework.
Multilevel algorithms repeatedly \emph{contract} vertices to obtain a hierarchy of \emph{coarser} hypergraphs while trying to preserve the cut structure.
On the coarsest hypergraph an \emph{initial partition} is computed in some way.
Then the contractions are reversed step-by-step and after every uncontraction a \emph{refinement} algorithm tries to improve the solution.
Most multilevel algorithms use a variant of the Fiduccia-Mattheyses (FM)~\cite{fm-a-82} or Kernighan-Lin (KL)~\cite{kl-efppg-70} local vertex moving heuristics.
These algorithms move vertices between blocks, prioritized by cut improvement.
The multilevel framework has been immensely successful because it provides a global view on the problem through local operations on the coarse levels.
Furthermore, it allows a great deal of engineering and tuning, which have a drastic impact on solution quality.
Even though this framework has been used since the 1990s, the implementations are still improving today.
A selection of well-known multilevel hypergraph partitioners are PaToH~\cite{patoh} (scientific computing), hMETIS~\cite{kaks-mhpav-99,kk-mkwhp-99} (VLSI design) KaHyPar~\cite{hs-icshp-17,hss-nfbrm-18} (general purpose, n-level), Zoltan~\cite{dbhbc-phpsc-06} (scientific computing, parallel), Zoltan-AlgD~\cite{scs-rbcmh-19} (algebraic distances based coarsening, sequential), Mondriaan~\cite{vb-atddd-05} (sparse matrices), MLPart~\cite{ahk-mcp-98} (circuit partitioning) and Par$k$way~\cite{tk-pmahp-08} (parallel).

Compared to graph partitioning, the performance of local vertex moving suffers from the presence of large hyperedges with vertices scattered over multiple blocks, since many moves have zero cut improvement.
On coarse levels of the multilevel hierarchy, this problem is alleviated since hyperedges contain fewer vertices.
A second remedy are flow-based refinement algorithms.
For graphs, Sanders and Schulz~\cite{ss-emgpa-11} extract a size-constrained corridor around the cut and compute a minimum cut within this corridor.
If the cut is balanced, an improved solution was found, otherwise the step is repeated with a smaller corridor.
Heuer~\etal~\cite{hss-nfbrm-18} extend their approach to hypergraphs by using \emph{Lawler networks}~\cite{l-cph-73}.
The Lawler network of a hypergraph is a flow network such that minimum $S$-$T$ hyperedge cuts can be computed via max-flow.

In their Flow-Balanced-Bipartition algorithm (FBB), Yang and Wong~\cite{yw-enfbm-96} use incremental maximum flows on the Lawler network to compute $\epsilon$-balanced hypergraph bipartitions.
Liu and Wong~\cite{hw-nfbmp-98} enhance FBB with a \emph{most-desirable-minimum-cut} heuristic, which is inspired by the correspondence between $S$-$T$ minimum cuts and closed node sets due to Picard and Queyranne~\cite{pq-osamc-82}.
It is similar to the \emph{most-balanced-minimum-cut} heuristics used in the multilevel graph partitioning tool KaHiP~\cite{ss-emgpa-11} and KaHyPar-MF~\cite{hss-nfbrm-18} as well as the \emph{avoid-augmenting-paths} piercing heuristics of FlowCutter~\cite{hs-gbpo-18} and HyperFlowCutter.
Li~\etal~\cite{llc-lvlsi-95} propose a push-relabel algorithm, which operates directly on the hypergraph.
Furthermore they present heuristics rooted in VLSI design for choosing sets of initial seed vertices $S$ and $T$ as well as piercing vertices.
The performance of their approach in other contexts than VLSI design remains unclear.

For perfectly balanced graph partitioning, diffusion-based methods have been successful~\cite{mms-a-09}.
Furthermore Sanders and Schulz~\cite{ss-tlagh-13} propose an algorithm based on detecting negative cycles, which is used on top of their evolutionary partitioner.
Delling and Werneck~\cite{dw-bbgb-12} provide an efficient implementation of an optimal branch-and-bound algorithm.
Additionally there are metaheuristic approaches such as PROBE~\cite{cbm-aprob-07}, as well as multilevel memetic algorithms due to Benlic and Hao~\cite{bh-aemma-10,bh-ammai-11,bh-aemts-11}.

\section{Preliminaries}\label{sec:preliminaries}
A \emph{hypergraph} $H=(V,E)$ consists of a set of $n$ vertices $V$ and a set of $m$ hyperedges $E$, where a hyperedge $e$ is a subset of the vertices $V$.
A vertex $v\in V$ is \emph{incident} to hyperedge $e \in E$ if $v \in e$.
The vertices incident to $e$ are called the \emph{pins} of $e$.
We denote the incident hyperedges of $v$ by $I(v)$ and its degree by $\deg(v):=|I(v)|$.
Furthermore let $p := \sum_{e \in E} |e|$ denote the total number of pins in $H$.
All hypergraphs in this paper are unweighted.
$H$ can be represented as a bipartite graph $G=(V \cup E, \{ (v,e) \in V \times E \mid v \in e\})$ with bipartite node set $V \cup E$ and an edge for every pin. 
This is also referred to as the \emph{star expansion} of $H$.
$H$ is \emph{connected} if its star expansion is connected.
Let $V[E'] := \bigcup_{e' \in E'} e'$ denote the vertex set induced by the hyperedge set $E'$.
To avoid confusion, we use the terms vertices, hyperedges and pins for hypergraphs, and we use the terms nodes and edges for graphs.

\subsection{Hypergraph Partitioning}
A \emph{bipartition} of a hypergraph $H$ is a partition $(A,B)$ of the vertices $V$ into two non-empty, disjoint sets (called blocks).
The \emph{cut} ${\cut(A,B) := \{ e \in E \mid e \cap A \neq \emptyset \wedge e \cap B \neq \emptyset \} }$ consists of all hyperedges with pins in both blocks.
The \emph{size} of a cut is the number of cut hyperedges $|\cut(A,B)|$.
Let $\epsilon \in [0,1)$. A bipartition $(A,B)$ is $\epsilon$-balanced if $\max(|A|,|B|) \leq \lceil (1+\epsilon)\frac{n}{2} \rceil$.
The \emph{balanced hypergraph bipartitioning problem} is to find an $\epsilon$-balanced bipartition $(A,B)$ which minimizes the cut. The special case $\epsilon = 0$ is called \emph{perfectly balanced bipartitioning}.

\subsection{Maximum Flows}
A flow network $\mathcal{N}=(\mathcal{V},\mathcal{E},S,T,c)$ is a simple symmetric directed graph $(\mathcal{V},\mathcal{E})$ with two non-empty \emph{terminal} node sets $S,T\subsetneq \mathcal{V}$, $S \cap T = \emptyset$, the source and target node set, as well as a capacity function $c : \mathcal{E} \mapsto \mathbb{R}_{\geq 0}$.
Any node that is not a source node and not a target node is a \emph{non-terminal} node.
A flow in $\mathcal{N}$ is a function $f:\mathcal{E} \mapsto \mathbb{R}$ subject to the \emph{capacity constraint} $f(e) \leq c(e)$ for all edges $e$, \emph{flow conservation} ${\sum_{(u,v)\in \mathcal{E}} f((u,v)) = 0}$ for all non-terminal nodes $v$ and \emph{skew symmetry} ${f((u,v))=-f((v,u))}$ for all edges~$(u,v)$. 
The \emph{value} of a flow ${|f| := \sum_{s \in S, (s,u)\in \mathcal{E}} f((s,u))}$ is the amount of flow leaving $S$.
The \emph{residual capacity} $r_f(e) := c(e) - f(e)$ is the additional amount of flow that can pass through $e$ without violating the capacity constraint.
The residual network with respect to $f$ is the directed graph $\mathcal{N}_f = (\mathcal{V},\mathcal{E}_f)$ where $\mathcal{E}_f := \{e \in \mathcal{E} | r_f(e) > 0\}$.
A node $v$ is \emph{source-reachable} if there is a path from $S$ to $v$ in $\mathcal{N}_f$, it is \emph{target-reachable} if there is a path from $v$ to $T$ in $\mathcal{N}_f$.
We denote the source-reachable and target-reachable nodes by $S_r$ and $T_r$, respectively.
An \emph{augmenting path} is an $S$-$T$ path in $\mathcal{N}_f$.
The flow $f$ is a \emph{maximum flow} if $|f|$ is maximal of all possible flows in $\mathcal{N}$. 
This is the case iff there is no augmenting path in $\mathcal{N}_f$.
An $S$-$T$ edge cut is a set of edges whose removal disconnects $S$ and $T$.
The value of a maximum flow equals the weight of a minimum-weight $S$-$T$ edge cut~\cite{ff-mftn-56}.
The \emph{source-side cut} consists of the edges from $S_r$ to $\mathcal{V} \setminus S_r$ and the \emph{target-side cut} consists of the edges from $T_r$ to $\mathcal{V} \setminus T_r$.
The bipartition $(S_r, \mathcal{V} \setminus S_r)$ is induced by the source-side cut and $(\mathcal{V} \setminus T_r, T_r)$ is induced by the target-side cut.
Note that $\mathcal{V} \setminus S_r \setminus T_r$ is not necessarily empty.

\subsection{Hyperedge Cuts via Maximum Flows}\label{sec:hg_cuts_via_flow}
Lawler~\cite{l-cph-73} uses maximum flows to compute minimum $S$-$T$ hyperedge cuts without balance constraints.
On the star expansion, the standard construction to model node capacities as edge capacities~\cite{amo-nf-93} is applied to the hyperedge-nodes.
A hyperedge $e$ is expanded into an \emph{in-node} $e_i$ and an \emph{out-node} $e_o$ joined by a directed \emph{bridge edge} $(e_i, e_o)$ with unit capacity.
For every pin $u \in e$ there are two directed \emph{external edges} $(u, e_i), (e_o, u)$ with infinite capacity.
The transformed graph is called the \emph{Lawler network}.
A minimum $S$-$T$ edge cut in the Lawler network consists only of bridge edges, which directly correspond to $S$-$T$ cut hyperedges in $H$.

Instead of using the Lawler network, we emulate max-flow algorithms directly on the hypergraph, using an approach first proposed by Pistorius and Minoux~\cite{pm-aidlm-03}.
In the paper it is formulated for unit weight hyperedges and the Edmonds-Karp flow algorithm~\cite{ek-tiaenf-72} but it can be easily extended to handle weighted hyperedges and emulate any flow algorithm.
For every hyperedge $e \in E$, we store the pins sending and receiving flow via $e$.
In this work, we consider only unit weight hyperedges and thus need to store only one pin $\flowfrom(e)$ sending flow into $e$, and one pin $\flowto(e)$ receiving flow from $e$.
To keep the description simple, it relies on this assumption as well.
Let $u$ be a fixed vertex.
The idea is to enumerate short paths of the form $u$ $\rightarrow e \in I(u) \rightarrow v \in e$ that correspond to paths in the residual Lawler network.
This allows us to emulate algorithms for traversing the residual Lawler network directly on the hypergraph, such as Breadth-First-Search or Depth-First-Search, as well as other types of local operations, \eg, the \emph{discharge} and \emph{push} operations in push-relabel algorithms~\cite{gt-namfp-88}.
For every hyperedge $e \in I(u)$ we do the following.
If $e$ has no flow, we enumerate all pins $v \in e$.
These paths correspond to $(u, e_i, e_o, v)$ in the Lawler network.
If $e$ has flow and $u = \flowto(e)$ we also enumerate all pins $v \in e$.
However, these paths correspond to $(u, e_o, e_i, v)$ in the Lawler network.
If $e$ has flow and $u = \flowfrom(e)$, there is no path in the residual Lawler network starting at $u$ that uses $e$.
If $e$ has flow and $\flowfrom(e) \neq u \neq \flowto(e)$, we enumerate just one path $(u,e,\flowfrom(e))$, corresponding to $(u, e_i, \flowfrom(e))$ in the Lawler network.
If we can push flow from the vertex $\flowfrom(e)$ to $T$, we can redirect the flow that the vertex $\flowfrom(e)$ sends into $e$, and instead route flow coming from $u$ to $\flowto(e)$.
Then $u$ becomes the new $\flowfrom(e)$.

We use this approach in our implementation because it is significantly more efficient than the Lawler network in practice.
In the last case we can avoid scanning all pins of $e$.
In a preliminary experiment, computing flow directly on the hypergraph yielded a speedup of 15 over using the Lawler network, for a hypergraph with maximum hyperedge size of only 36.
The speedup will be more extreme on hypergraphs with larger hyperedges.

Via the Lawler network and the above emulation approach, the notions of flow, source-reachable vertices and source-side cuts translate naturally from graphs to hypergraphs.
We use the notation and terminology already known from max-flows in graphs.

\section{HyperFlowCutter}\label{sec:hfc}
In the following we outline the core HyperFlowCutter algorithm, which can only be used on connected hypergraphs.
Then we discuss how to handle disconnected hypergraphs, and finally show how to improve an existing partition using HyperFlowCutter.
\subsection{The Core Algorithm}\label{sec:core}
\begin{figure}
	\begin{subfigure}[t]{.32\linewidth}
		\includegraphics[width=\linewidth]{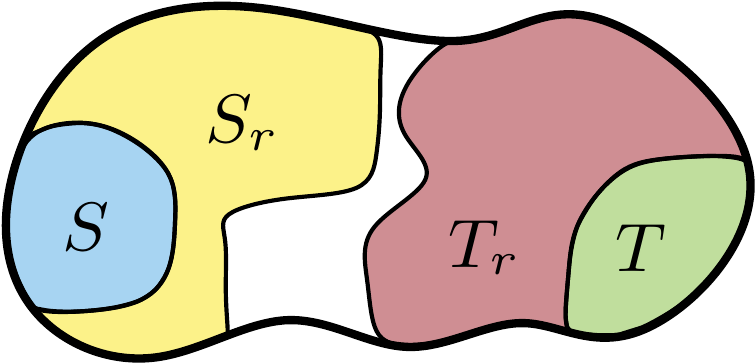}
		\caption{Compute minimum $S$-$T$ cuts.}\label{fig:fc_initialstate}
	\end{subfigure}
	\hfill
	\begin{subfigure}[t]{.32\linewidth}
		\includegraphics[width=\linewidth]{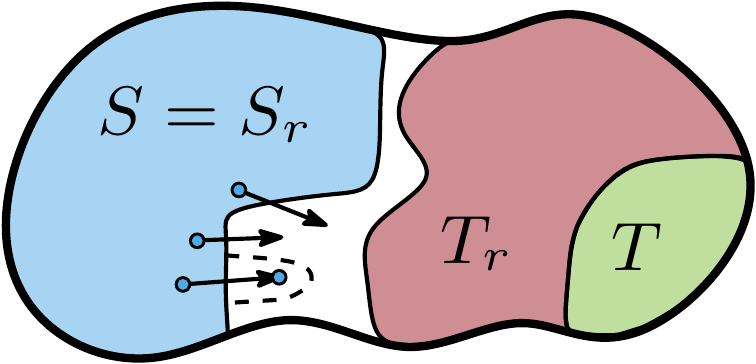}
		\caption{Add $S_r$ to $S$ and choose a piercing vertex.}\label{fig:fc_pierce}
	\end{subfigure}
	\hfill
	\begin{subfigure}[t]{.32\linewidth}
		\includegraphics[width=\linewidth]{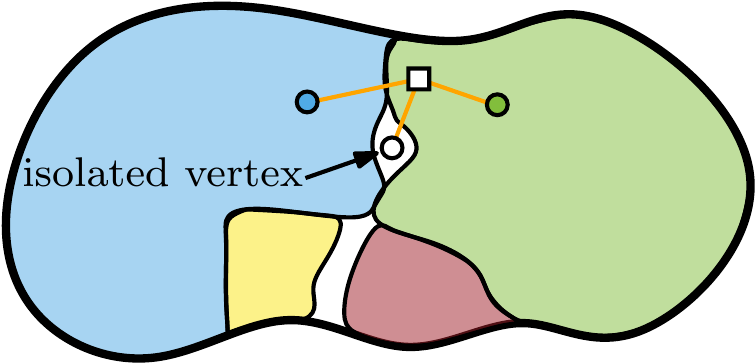}
		\caption{Mixed hyperedge (white square) with incidence relations (orange) and an isolated vertex (white disk).}\label{fig:fc_isolated}
	\end{subfigure}
	\caption{Flow augmentation and computing $S_r, T_r$ in Fig.\ref{fig:fc_initialstate}; adding $S_r$ to $S$ and piercing the source-side cut in Fig.\ref{fig:fc_pierce}. $S$ in blue, $S_r \setminus S$ in yellow, $T$ in green, $T_r\setminus T$ in red, $V \setminus S_r, \setminus T_r$ in white.}\label{fig:illuHFC}
\end{figure}

The idea of the \emph{core} HyperFlowCutter algorithm is to solve a sequence of incremental $S$-$T$ max-flow min-cut problems with monotonically increasing cut size and balance, until an $\epsilon$-balanced bipartition is found.
They are incremental in the sense that the terminals $S,T$ of the current flow problem are subsets of the terminals in the next flow problem, which allows us to reuse previously computed flows.

Given starting terminal sets $S_\text{init},T_\text{init}$, we set $S:= S_\text{init}$, $T := T_\text{init}$
First, we compute a maximum $S$-$T$ flow.
We terminate if the bipartition $(S_r, V \setminus S_r)$ induced by the source-side cut or $(V \setminus T_r, T_r)$ induced by the target-side cut is $\epsilon$-balanced.
Otherwise, we add the source-reachable vertices $S_r$ to $S$, if $|S_r| \leq |T_r|$, or we add $T_r$ to $T$ if $|S_r| > |T_r|$.
Assume $|S_r| \leq |T_r|$ without loss of generality.
Further, we add one or more vertices, called \emph{piercing vertices}, to $S$.
This step is called \emph{piercing} the cut.
It ensures that the next flow problem yields a different bipartition.
Subsequently, we augment the previous flow to a max-flow that respects the new sources.
We repeat these steps until an $\epsilon$-balanced bipartition is found.

Note that after adding $S_r$ to $S$, the flow is still a maximum $S$-$T$ flow, even though the added vertices are now exempt from flow conservation.
Using the smaller side allows it to catch up with the larger side.
In particular, this ensures that $\epsilon$-balance is always possible, as neither side grows beyond $\lceil n/2 \rceil$ vertices.

A hyperedge with pins in both $S$ and $T$ is \emph{mixed}, all other hyperedges are \emph{non-mixed}.
We consider two options to find piercing vertices.
The preferred option is to choose all pins $e \setminus S \setminus T$ of a non-mixed hyperedge $e$ in the source-side cut.
Adding multiple vertices is faster in practice.
This small detail is a major difference to FBB~\cite{yw-enfbm-96} and is necessary to make the running time feasible on the used benchmark set.
If the source-side cut contains only mixed hyperedges, we choose a single non-terminal vertex incident to the source-side cut.
We prefer piercing vertices which are not reachable from $T$, as these avoid augmenting paths in the next iteration, and thus the cut size does not increase.
Avoiding augmenting paths has the highest precedence, followed by choosing hyperedges over single vertices.
Ties are broken randomly.
If the piercing vertices are not reachable from $T$, we do not recompute $T_r$ and we skip flow augmentation, but we do recompute $S_r$.

We experimented with other piercing methods, including trying to avoid mixed hyperedges, the distance-based piercing of FlowCutter, as well as piercing based on a machine learning technique named ensemble classification.
We discuss ensemble classification again in the experimental section, although in a different context.
None of these approaches yielded consistent or significant quality improvements over just avoiding augmenting paths and random tie-breaking, which is why we use only those.

\subparagraph{Asymptotic Complexity.}
The asymptotic running time of Core HyperFlowCutter is $\mathcal{O}(cp)$ where $c$ is the cut size of the $\epsilon$-balanced partition and $p$ is the number of pins in the hypergraph.
Roughly speaking, the term $p$ stems from performing up to one hypergraph traversal per flow unit.
Here we use that $n\leq p, m \leq p$ holds for connected hypergraphs.
Note that the running time is output-sensitive, however the factor $c$ is rather pessimistic in practice, since the flow algorithm finds many augmenting paths in a single traversal.
The original proof for FlowCutter~\cite{hs-gbpo-18} is applicable, but implementing the piercing step requires a little care.
Selecting piercing vertices takes $\mathcal{O}(c)$ per iteration, and we have at most $n \leq p$ iterations.
Selecting a non-mixed hyperedge for piercing takes $\mathcal{O}(c)$ time, by iterating over the cut hyperedges.
Selecting single piercing vertices which avoid augmenting paths whenever possible, is slightly more involved, when restricted to $\mathcal{O}(c)$ time.
For every cut hyperedge $e$ we additionally track its pins that are not reachable from either side.
This can be implemented with a linked list, from which we delete vertices once they get reachable from a side.
An alternative implementation divides the memory storing the pins of $e$ into three regions: the pins in $S_r$, in $T_r$ or not reachable.
Then we can check, whether $e$ has pins that are not reachable from either side, and if so pick one.
In practice, this adds significantly to the complexity of the implementation and the piercing step is never critical for running time, so our implementation simply scans all non-terminal boundary vertices.

Our implementation has $\mathcal{O}(n+m)$ memory footprint by transferring and re-engineering the implementation details from~\cite{hs-gbpo-18}.
This is dominated by the $\mathcal{O}(n+m+p)$ memory for storing the hypergraph.

\subparagraph{Isolated Vertices.}
We call a vertex $v \notin S \cup T$ \emph{isolated} if every incident hyperedge $e \in I(v)$ is mixed.
Figure~\ref{fig:fc_isolated} illustrates an isolated vertex.
An isolated vertex cannot be reached from either $S$ or $T$ via hyperedges not in the cut.
Mixed hyperedges remain in both the source-side cut and the target-side cut.
Thus isolated vertices can be moved freely between the two blocks to increase balance.
It never makes sense to permanently add them to a side, so we exclude them from the piercing step.
Furthermore, this needs to be reflected when checking for $\epsilon$-balance.
For checking the bipartition $(S_r, V \setminus S_r)$ we assume up to $n/2 - |S_r|$ of the isolated vertices were part of $S_r$, analogously for $T_r$.

\subparagraph{Maximum Flow Algorithm.}\label{sec:maxflowalgo}
In our implementation, we adapt Dinic maximum flow algorithm~\cite{d-aspmf-70} to operate directly on the hypergraph, as described in Section~\ref{sec:hg_cuts_via_flow}.
Dinic algorithm has two alternating phases that are repeated until a maximum flow is found: computing hop distance labels of nodes, using Breadth-First-Search, and finding augmenting paths using Depth-First-Search, such that the distance labels always increase by one along the path.
We interleave the first phase of Dinic algorithm with the computation of reachable vertices, in order to avoid duplicate hypergraph traversal.
This intrusive optimization is important for improving the running time of flow augmentation in practice, as the part of the running time of the flow algorithm that cannot be charged towards computing reachable vertices is dominated by the part that can be.
This is not possible with push-relabel algorithms~\cite{gt-namfp-88}, which is why they were ruled out after short experimentation.
We experimented with the Edmonds-Karp flow algorithm~\cite{ek-tiaenf-72}, modified to augment flow along multiple vertex-disjoint paths in one Breadth-First-Search by propagating vertex labels through the layers.
It is slightly slower than Dinic for plain HyperFlowCutter but unfeasible for the refinement variant of HyperFlowCutter, since there are fewer vertices to route flow through and thus the amount of flow augmented per Breadth-First-Search is limited by a few bottleneck vertices.

\subsection{Disconnected Hypergraphs}\label{sec:disconnectedhypergraphs}
The HyperFlowCutter core algorithm is limited to connected hypergraphs since it computes $S$-$T$-cuts.
An obvious approach for handling disconnected hypergraphs is connecting components artifically.
We refrained from this because a component that intersects neither $S$ nor $T$ would be added to $S$ or $T$ in its entirety.
Instead, we run the core algorithm up to $\epsilon = 0$ on every connected component.
The core algorithm computes multiple bipartitions with different balances.
We systematically try all possible ways to combine the bipartitions of the components into a bipartition of $H$.

This can be stated as a generalization of the well-known \textsc{SubsetSum} problem.
\textsc{SubsetSum} is a weakly NP-hard decision problem, which asks whether a subset of an input multiset of positive integers $A=\{a_1,\dots, a_z\}$, the \emph{items}, sums to an input target sum $W$.
Finding a bipartition with zero cut is equivalent to \textsc{SubsetSum}, where the items are the sizes of the components and $W$ is the minimum size of the smaller block.
We are interested in any subset summing to at least $W$.
Let $A$ be sorted in increasing order and let $Q(i,S)$ be a boolean variable, which is true iff a subset of the first $i$ items sums to $S$.
The standard pseudo-polynomial time dynamic program (DP)~\cite[Section 35.5]{clrs-ia-01} for \textsc{SubsetSum} computes solutions for all possible target sums.
It fills the DP table $Q$ by iterating through the items in increasing order and setting $Q(i,S)$ to true if $Q(i-1,S-a_i)$ or $Q(i-1,S)$ is true.
For filling row $i$, only row $i-1$ is required, so the memory footprint is not quadratic.

We now turn to non-zero cut bipartitions by allowing to split items in different ways and associating costs with the splits.
The core algorithm computes multiple bipartitions $P_i$ on component $C_i$, at most one for every possible size of the smaller side.
These correspond directly to the different ways we can split items.
The associated cost is the cut size.
We modify the standard \textsc{SubsetSum} DP to minimize the added cuts instead of finding any subset, ensuring every component/item is split only one way in a solution, \ie, select a bipartition, and trying the smaller and larger side of the component bipartition for the smaller side of the bipartition of $H$.
The worst case asymptotic running time of this DP is $\mathcal{O}(\sum_{i=1}^z \sum_{j=1}^{i-1} |P_i||P_j|)$.

We propose some optimizations to make the approach faster in practice.
First we solve standard \textsc{SubsetSum} to check whether $H$ has an $\epsilon$-balanced bipartition with zero cut.
For the \emph{gap-filler} optimization, we find the largest $g \in \mathbb{N}$ such that for every $x \in [0,g]$ there are connected components, whose sizes sum to $x$.
Computing $g$ is possible in $\mathcal{O}(n)$ time.
Let $C_1, \dots, C_z$ be sorted by increasing size, which takes $\mathcal{O}(n)$ time using counting sort~\cite[Section 8.2]{clrs-ia-01}.
Then $g=\sum_{j=1}^{k-1} |C_j|$ for the smallest $k$ such that $|C_k| > 1 + \sum_{j=1}^{k-1} |C_j|$.
It is never beneficial to split the components $C_1, \dots, C_{k-1}$.
For most hypergraphs we consider in our experiments, we do not invoke the DP because, due to gap-filler, we split only the largest component.
For the hypergraphs on which we do invoke the DP, its running time is negligible.
Nonetheless, it is easy to construct a worst case instance, where the quadratic running time is prohibitive.
For a robust algorithm, we propose to sample bipartitions from every $P_i$ so that the worst case running time falls below some input threshold.
The samples should include balanced bipartitions to guarantee that a balanced partition on $H$ can be combined from those on $H[C_i]$.

\subsection{HyperFlowCutter as a Refinement and Balancing Algorithm}\label{sec:refinement}
Instead of partitioning from scratch, HyperFlowCutter can be used to refine an existing balanced bipartition $\pi = (V_1, V_2)$, or repair the balance of an unbalanced bipartition.
We fix two blocks of vertices $V'_i \subset V_i$ such that $|V'_i| \leq \alpha \cdot n$ for a \emph{relative block-size threshold} parameter $\alpha \in [0, 0.5]$.
To obtain $V'_i$ we run Breadth-First-Search from the boundary vertices on the side of $V_i$ 
until $|V_i|-\alpha \cdot n$ vertices have been visited.
The $\alpha \cdot n$ vertices not visited by the Breadth-First-Search are set as $V'_i$.
Then we run HyperFlowCutter with $S=V'_1, T=V'_2$.
We call this algorithm \emph{ReBaHFC}.
This idea is equivalent to the way KaHyPar and KaHiP extract \emph{corridors} around the cut for their flow-based refinement.
Only the semantics of the size constraint are different to our approach.
However, KaHyPar and KaHiP only compute one flow.
If the associated bipartition is not balanced, a smaller flow network is derived.
This is repeated until the bipartition is balanced.
ReBaHFC does not need to repeatedly re-scale flow networks.

In this work we only perform refinement as a post-processing step to a given partition, whereas KaHyPar and KaHiP employ flow-based refinement on the multilevel hierarchy.
In a future work we hope to integrate HyperFlowCutter based refinement into KaHyPar.
Using ReBaHFC could eliminate the significant overhead of repeated rescaling and improve solution quality -- in particular when the minimum cut is just short of being balanced.

We use the fast multilevel partitioner PaToH~\cite{patoh} to obtain initial partitions.
We briefly discuss properties of PaToH and differences between its presets.
For coarsening, PaToH uses agglomerative clustering, based on the number of common hyperedges divided by cluster size.
For initial partitioning, a portfolio of graph growing, bin packing and neighborhood expansion algorithms is used.
For refinement PaToH uses a pass of FM~\cite{fm-a-82} followed by a pass of KL~\cite{kl-efppg-70}, each initialized with boundary nodes.

In order to improve cut size, Walshaw~\cite{w-mrcop-04} proposed to iterate the multilevel scheme by contracting only nodes from the same block, which maintains the cut size, thus allowing refinement on coarse levels starting from a relatively high quality partition.
The existing partition serves as initial partition on the coarsest level.
One iteration is called a V-cycle.
Contraction can be stopped at different stages.
The quality preset PaToH-Q, uses 3 full V-cycles and 3 shorter V-cycles as opposed to the single V-cycle of the default preset PaToH-D.
To accelerate partitioning, both presets temporarily discard hyperedges which have more pins than some threshold.
PaToH-D sets a lower threshold than PaToH-Q.

\section{Experimental Evaluation}\label{sec:experiments}
We implemented HyperFlowCutter and ReBaHFC in C++17 and compiled our code using g++8 with flags \texttt{-O3 -mtune=native -march=native}.
The source code is available on GitHub\footnote{Souce code available at~\url{https://github.com/kit-algo/HyperFlowCutter}}.
Experiments are performed sequentially on a cluster of Intel Xeon E5-2670 (Sandy Bridge) nodes with two Octa-Core processors clocked at 2.6 GHz with 64 GB RAM, 20~MB L3- and 8$\times$256 KB L2-Cache, using only one core of a node.

We use the benchmark set\footnote{Benchmark set and detailed statistics available at \url{https://algo2.iti.kit.edu/schlag/sea2017/}} of Heuer and Schlag~\cite{hs-icshp-17}, which has been used to evaluate KaHyPar.
It consists of 488 hypergraphs from four sources: the ISPD98 VLSI Circuit Benchmark Suite~\cite{a-tispd-98} (VLSI, 18 hypergraphs), the DAC 2012 Routability-Driven Placement Benchmark Suite~\cite{naslw-tdacr-12} (DAC, 10), the SuiteSparse Matrix Collection~\cite{dh-tufsm-11} (SPM, 184) and the international SAT Competition 2014~\cite{sat2014} (Literal, Primal, Dual, 92 hypergraphs each).
The set contains 173 disconnected hypergraphs, in particular all DAC instances are disconnected.
Refer to~\cite{hs-icshp-17} for more information on how the hypergraphs were derived.
Unless mentioned otherwise, experiments are performed on the full benchmark set.

In the following we describe the configuration of ReBaHFC and plain HyperFlowCutter, before comparing them to competing algorithms in Section~\ref{sec:comparison}.

\subsection{General HyperFlowCutter Configuration}\label{sec:parameterstudy}

\begin{table}[tb]
	\centering
	\caption{Average and quantile speedups of the hybrid and interleaved execution strategies.}\label{table:execution_strategy}
	\begin{tabular}{l *{8}{r}}
		\toprule
		& avg & min &  0.1 & 0.25 & median & 0.75 & 0.9 & max \\ \midrule
		hybrid & 2.21 & 0.4 & 0.96 & 1.07 & 1.29 & 1.55 & 2.57 & 49.66 \\
		interleaved & 3.88 & 0.55 & 1.0 & 1.14 & 1.38 & 1.74 & 2.66 & 175.47 \\ 
		\bottomrule
	\end{tabular}
\end{table}

To improve the solution quality of HyperFlowCutter, we run it $q \in \mathbb{N}$ times with different terminal pairs and take the minimum cut.
To improve running time we run them simultaneously, in an \emph{interleaved} fashion, as already described in~\cite{hs-gbpo-18}, so that the output-sensitive running time depends on the smallest found $\epsilon$-balanced cut, not the largest.
We always schedule the terminal pair with the currently smallest cut to progress.

Table~\ref{table:execution_strategy} shows the average and some quantile speedups when interleaving the execution of 20 random terminal vertex pairs, instead of running them one after another; repeated for 5 random seeds.
Because consecutive execution exhibits more memory locality, we also tested a hybrid strategy where the instance with the currently smallest cut is allowed to make multiple progress iterations.
Interleaving outperforms consecutive execution by a factor of 3.88 on average and also consistently beats hybrid execution.
This shows that saving work is more important than memory locality.
These numbers stem from a preliminary experiment on a 139 hypergraph subset of the full benchmark set.
The subset contains 78 sparse matrices, 17 Primal, 23 Dual, 6 Literal SAT instances, 15 VLSI and 0 DAC instances.
It contains only connected hypergraphs (all DAC instances are disconnected) in order to measure the impact of interleaving, not the setup overhead for many small connected components.

\subsection{ReBaHFC Configuration}\label{sec:config:rebahfc}
\begin{figure}[tb]
	\centering
	\includegraphics[width=0.6\linewidth]{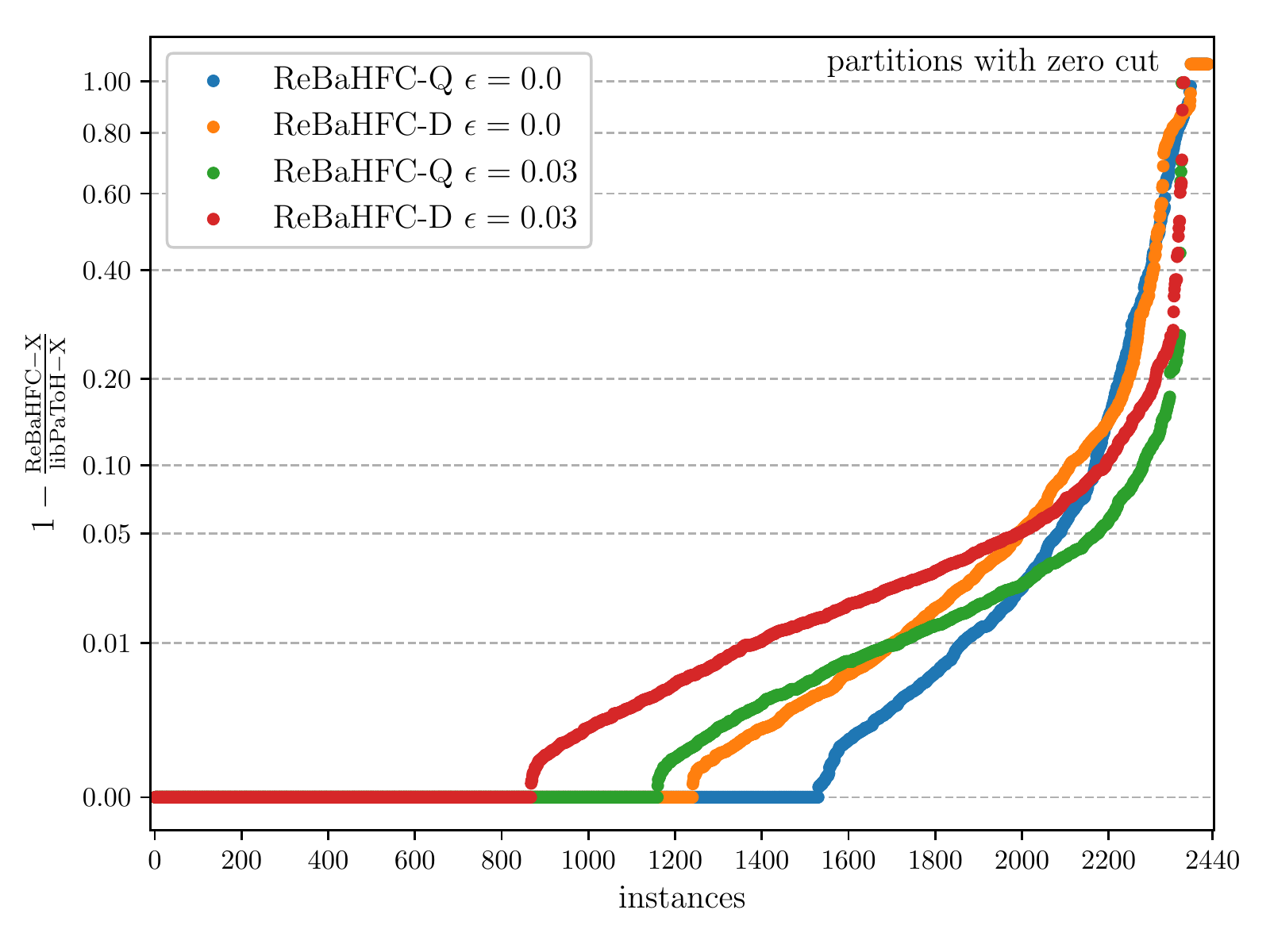}
	\caption{Improvement ratios $1 - \frac{\cut(\mathrm{ReBaHFC})}{\cut(\mathrm{PaToH})}$ of the two ReBaHFC variants and $\epsilon = 0, 0.03$ compared to their initial partitions. The curves are independent from one another. Higher values are better. Ratios of zero mean ReBaHFC did not improve the initial partition. We count partitions with zero cut as an improvement ratio greater than $1$, since ReBaHFC does not invoke PaToH in this case.} 
	\label{fig:experimental:rebahfc:improvementratios}
\end{figure}

We now discuss the configuration for ReBaHFC.
The imbalance for the initial partition is set to the same value as the desired imbalance $\epsilon$ for the output partition, which proved superior to larger imbalances on initial partitions.
The block-size threshold parameter $\alpha$ should depend on $\epsilon$, so we settled on $\alpha = 0.4$ for $\epsilon = 0.03$ and $\alpha = 0.46$ for $\epsilon = 0$.
We resize the blocks once and run HyperFlowCutter five times, interleaved as described in the previous section.
This number seems to provide decent quality without increasing running time too much.
We consider two variants: ReBaHFC-D, which uses PaToH with default preset and ReBaHFC-Q, which uses PaToH with quality preset.
The parameter study, which led to these choices, is discussed in more detail in Appendix~\ref{sec:experimental:config_rebahfc}.

Figure~\ref{fig:experimental:rebahfc:improvementratios} shows how much ReBaHFC improves the initial partition.
We run the two ReBaHFC variants for $\epsilon = 0, 0.03$ on all hypergraphs of the benchmark set with five different random seeds and plot the ratio $1 - \frac{\cut(\mathrm{ReBaHFC})}{\cut(\mathrm{PaToH})}$ per run.
Note that there is no comparison between the curves, and higher values are better for ReBaHFC.
Table~\ref{table:rebahfc_impr_by_category} reports how often ReBaHFC improves the initial partition, for different hypergraph classes.
As expected, ReBaHFC-Q could improve fewer solutions than ReBaHFC-D since the PaToH baseline is already better.
Furthermore, ReBaHFC has more opportunities for refinement with $\epsilon = 0.03$, in particular on the DAC and VLSI instances, whereas it struggles with the Primal and Literal SAT instances for $\epsilon = 0$.
Note that other refinement algorithms do not always improve solutions either.
In particular, local moving based refinement algorithms struggle with zero-gain moves in the presence of large hyperedges, and the flow-based refinement in KaHyPar can yield unbalanced solutions or reproduce the existing solution.
These results show that HyperFlowCutter is a promising candidate for a refinement algorithm integrated in a multilevel partitioner, which is a direction we hope to investigate in future work.

The PaToH runs in the experiments from Section~\ref{sec:comparison} use other random seeds than those used internally in ReBaHFC.
This makes sure that stand-alone PaToH can find smaller cuts than ReBaHFC.

\begin{table}[tb]
	\centering
	\caption{Overview by hypergraph class, how often ReBaHFC improves the initial partition.}
	\label{table:rebahfc_impr_by_category}
	\begin{tabular}{lrrrrrrrr}
\toprule
      Algorithm &  $\epsilon$ &  all &  SPM & Dual & Primal &  Lit &  DAC & VLSI \\
\midrule
 ReBaHFC-Q &        0.00 & 37.3 & 39.8 & 47.4 &   23.9 & 33.0 & 66.0 & 34.4 \\
 ReBaHFC-D &        0.00 & 49.2 & 58.4 & 59.8 &   27.4 & 41.3 & 58.0 & 47.8 \\
 ReBaHFC-Q &        0.03 & 52.5 & 47.2 & 51.5 &   50.9 & 57.8 & 82.0 & 76.7 \\
 ReBaHFC-D &        0.03 & 64.5 & 61.5 & 65.9 &   56.7 & 71.1 & 76.0 & 86.7 \\
\bottomrule
\end{tabular}

\end{table}

\subsection{Plain HyperFlowCutter Configuration}\label{sec:config:plain_hfc}

For the experiments on perfectly balanced partitioning we run plain HyperFlowCutter with up to $q=100$ terminal pairs and take the minimum cut.
This value was used already for FlowCutter~\cite{hs-gbpo-18}.
With plain HyperFlowCutter we want to push the envelope on solution quality for $\epsilon = 0$, regardless of running time -- because, as the experiments show, ReBaHFC already provides a good time-quality trade-off.
The most simple method for choosing starting terminals is to select random vertices.
We unsuccessfully experimented with \emph{pseudo-peripheral} terminals, \ie two vertices that are intuitively far away from each other and at the boundary of the hypergraph.
Instead we propose a selection method based on ensemble classification.
Ensemble classification is a technique used in machine learning to build a strong classifier from multiple weak ones.
We compute $10$ bipartitions $\pi_1, \dots, \pi_{10}$ with PaToH-D.
Let $x \equiv y  \Leftrightarrow \pi_i(x)=\pi_i(y)$ for all $i=1,\dots,10$ be the equivalence relation, in which two vertices are equivalent if they are in the same block for all ensemble bipartitions.
An equivalence class is likely in the same block of a good bipartition and is thus suited as a terminal set.
We order the equivalence classes by size in descending order and group two successive classes as one terminal pair.
Generally speaking, the larger equivalence classes make for better terminal pairs.
Based on experiments in Appendix~\ref{sec:experimental:config_plain}, we use 3 ensemble terminal pairs and 97 random vertex pairs.
The reported running time for plain HyperFlowCutter always includes the running time for the 10 PaToH-D runs.

On 42 of the 488 hypergraphs, plain HyperFlowCutter with $100$ terminal pairs exceeds the eight hour time limit.
One downside of interleaving executions is that the solution is only available once all terminal pairs have been processed.
Instead of interleaving all 100 executions, we run four waves of $\langle 1,5,14,80 \rangle¸$ terminal pairs consecutively and interleave execution within waves.
An improved bipartition is available after every wave, so that, even if the time limit is exceeded, a solution is available as long as the first wave has been completed.
We chose wave sizes, so that completing waves four and three corresponds to 100 and 20 terminal pairs, respectively, as these values were used in~\cite{hs-gbpo-18}.
The first wave consists of the first ensemble terminal pair, the second/third wave consist of 5/14 random terminal pairs and the fourth wave consists of 78 random as well as two additional ensemble terminal pairs.
There are 438 hypergraphs for which the fourth wave finishes, 35 for which the third but not the fourth wave finishes, 6 for the second, 1 for the first and there are 8 hypergraphs which are partitioned with zero cut, using just the subset sum preprocessing.

\subsection{Comparing ReBaHFC and HyperFlowCutter against State-of-the-Art Hypergraph Partitioners}\label{sec:comparison}
\begin{figure}[tb]
	\begin{subfigure}[t]{.49\linewidth}
		\includegraphics[width=\linewidth]{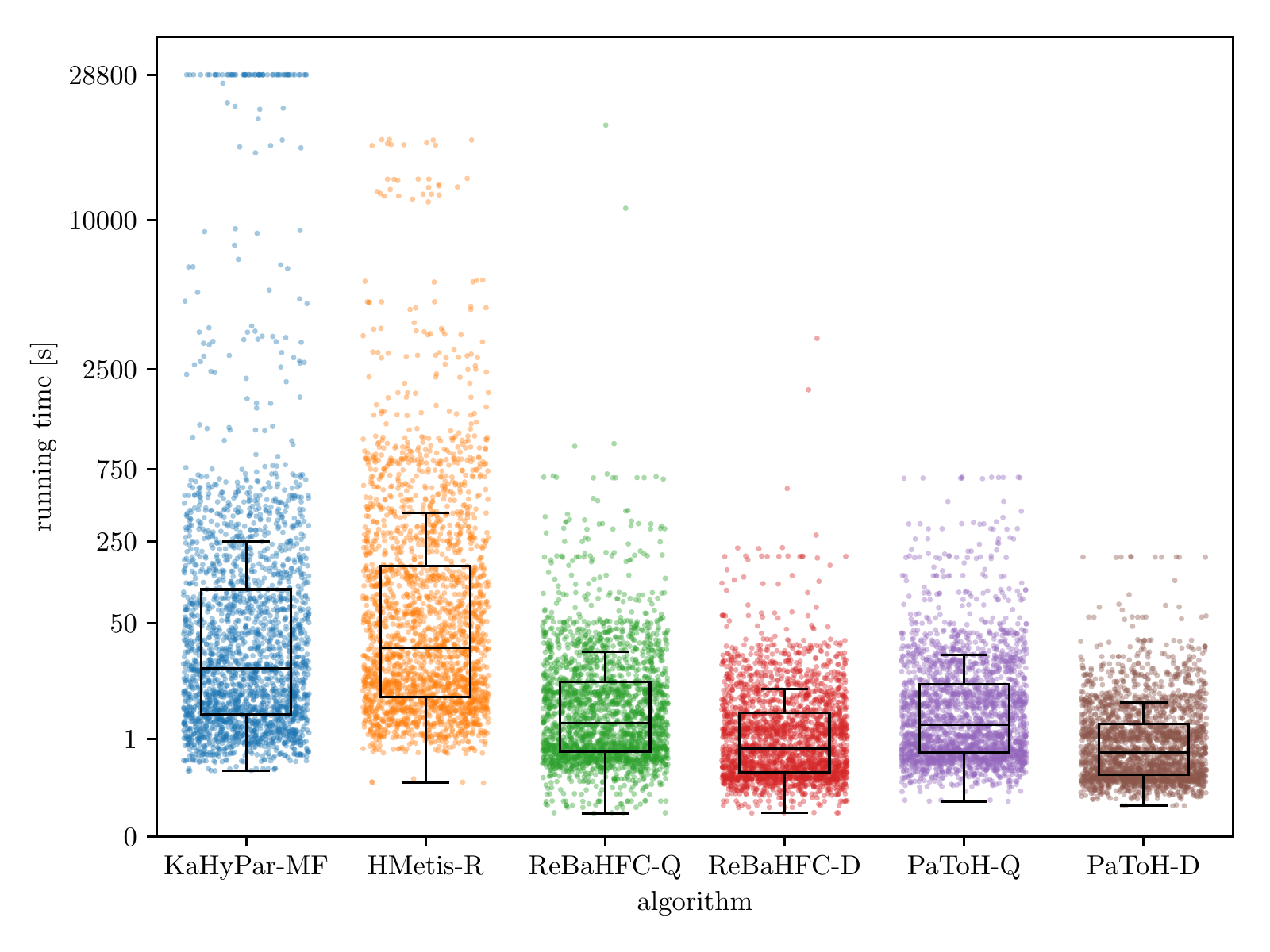}
	\end{subfigure}
	\hfill
	\begin{subfigure}[t]{.49\linewidth}
		\includegraphics[width=\linewidth]{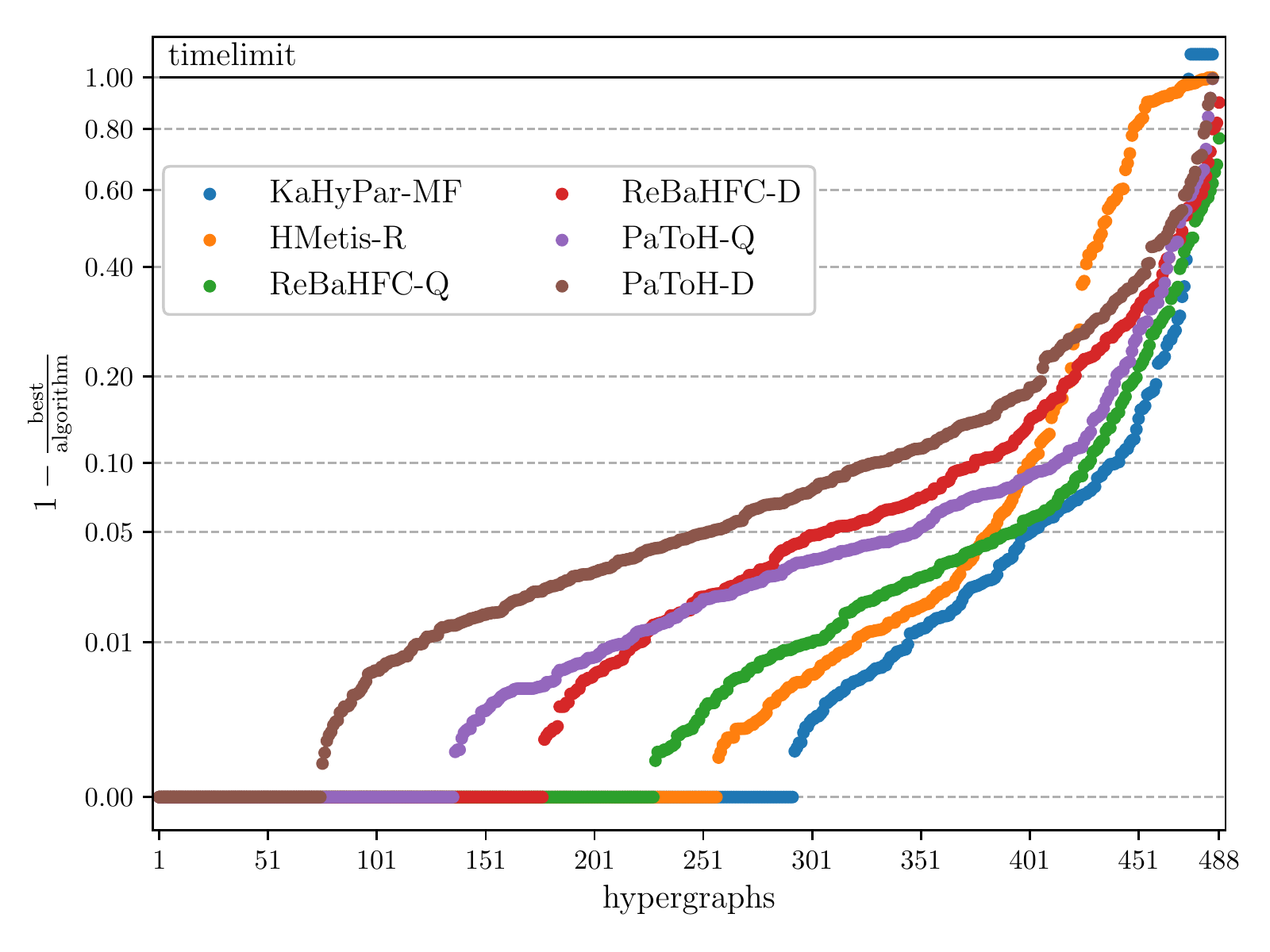}
	\end{subfigure}
	\caption{Comparison between the algorithms for $\epsilon = 0.03$. Left: Absolute running times for every hypergraph and random seed. Right: Performance plot relating the minimum cut per algorithm and hypergraph to the overall best cut for that hypergraph. Lower values are better.}
	\label{fig:experimental:eps0.03}
\end{figure}

In this section, we compare ReBaHFC and plain HyperFlowCutter against state-of-the-art hypergraph partitioners.
After discussing our comparison methodology, we present results for two settings, namely $\epsilon = 0.03$ and $\epsilon = 0$.

\subparagraph{Methodology.}
We run each partitioner five times with different random seeds and report the minimum cut.
For every run we set a time limit of eight hours.
We use the \emph{performance plots} introduced in~\cite{shhmss-hplrb-16} to compare algorithms on a per-hypergraph basis regarding cut size.
For each algorithm and hypergraph these plots contain a \emph{performance ratio} $1 - \text{best}/\text{algorithm}$, which relates the minimum cut found by any algorithm to the minimum cut found by this algorithm.
The ratios of each algorithm are sorted in increasing order.
A ratio of $0$ means that this algorithm found the smallest overall cut, the number of achieved ratios of $0$ is the number of hypergraphs on which this algorithm is the best.
Furthermore, algorithm A dominates algorithm B if the curve of A is strictly below that of B.
We use values greater than $1$ to indicate that algorithms exceeded the time limit or produced unbalanced solutions.
This is clearly marked in the plots.
To include partitions with zero cut, we set the performance ratio to $0$, if the algorithm found the zero cut partition, and $1$ otherwise.
The performance plots use a cube root scaled y-axis in order to reduce right skewness~\cite{cuberoots} and to give a fine-grained view on the smaller improvements.
For comparing algorithms regarding running time we use a combination of a scatter plot, which shows every measured running time, and a box plot (0.25, median, 0.75 quantiles, whiskers at most extreme points within distance $1.5 \cdot \operatorname{IQR}$ from the upper/lower quartile).
The running time plots use a fifth root scaled y-axis for a fine-grained view on areas with smaller running times, which contain more data points.

\subparagraph{Comparison for 3\% imbalance.}
For $\epsilon = 0.03$ we compare ReBaHFC against the state-of-the-art hypergraph partitioning tools KaHyPar-MF (the latest version of KaHyPar with flow-based refinement) and hMETIS-R (the recursive bisection variant of hMETIS), as well as PaToH-D (default preset) and PaToH-Q (quality preset).
We use the library interface of PaToH.
According to the hMETIS manual, hMETIS-R is preferred over hMETIS-K (direct k-way) for bipartitions, so we exclude hMETIS-K.
These tools were chosen because they provide the best solution quality according to~\cite{ahss-ehpa-17, hs-icshp-17}.
We chose $\epsilon = 0.03$ as this is a commonly used value in the literature.
Plain HyperFlowCutter is excluded from this part of the experiments because it is not competitive.

Figure~\ref{fig:experimental:eps0.03} shows the running times and a performance plot on the full benchmark set for $\epsilon = 0.03$.
In addition to the running time plot, we compare algorithms by the geometric mean of their running times.
We use the geometric mean in order to give instances of different sizes a comparable influence.
KaHyPar-MF finds the smallest cut on 292 hypergraphs, hMETIS-R on 257, ReBaHFC-Q on 228, ReBaHFC-D on 177, PaToH-Q on 136 and PaToH-D on 75 of the 488 hypergraphs.
While KaHyPar-MF is the best algorithm regarding solution quality, it is also the slowest, exceeding the time limit on 11 hypergraphs.
For the instances on which ReBaHFC-Q does not find the best solution it provides solution quality similar to hMETIS-R and only marginally worse than KaHyPar-MF.
In particular its solution quality compared to the best cut deteriorates less than that of hMETIS-R.
With 2.23s PaToH-Q is one order of magnitude faster than KaHyPar (34.1s) and hMETIS-R (20.1s), whereas ReBaHFC-Q (2.32s) is only slightly slower than PaToH-Q.
Furthermore ReBaHFC-D (0.68s) finds more of the best solutions than PaToH-Q at a running time between PaToH-D (0.5s) and PaToH-Q.
Thus ReBaHFC-Q and ReBaHFC-D provide new Pareto points in the time-quality trade-off.
In Appendix~\ref{sec:experimental:performanceplots_instanceclasses} Figure~\ref{fig:experimental:performanceplots_instanceclasses} shows performance plots for the different hypergraph classes of the benchmark set.
ReBaHFC is particularly good on the DAC and SPM instances.
There are hypergraphs on which ReBaHFC is faster than PaToH.
These are disconnected hypergraphs, for which ReBaHFC invokes PaToH on smaller sub-hypergraphs, due to the gap-filler optimization and the \textsc{SubsetSum} preprocessing described in Section~\ref{sec:disconnectedhypergraphs}.
	
\begin{figure}[tb]
	\begin{subfigure}[t]{.49\linewidth}
		\includegraphics[width=\linewidth]{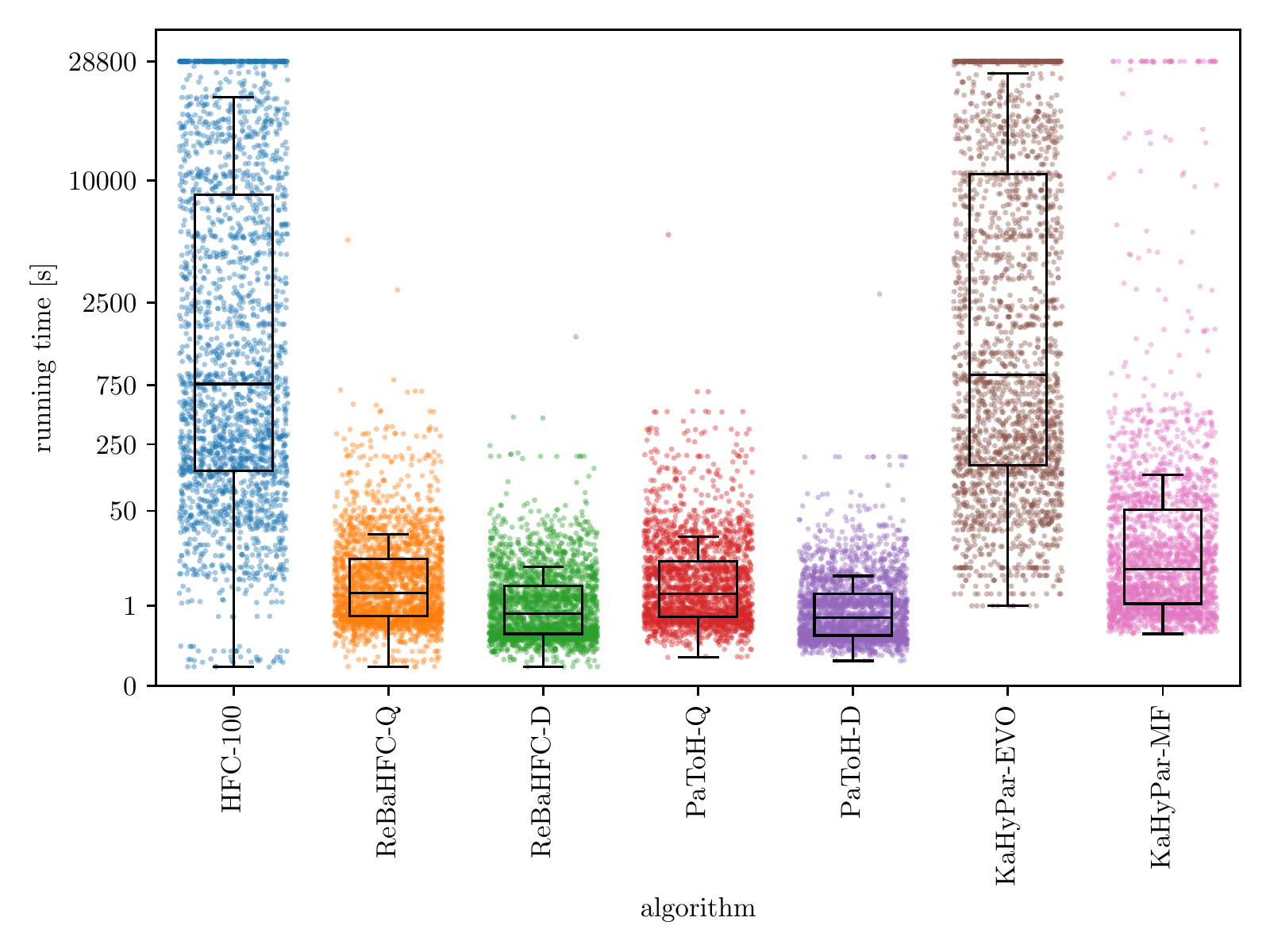}
	\end{subfigure}
	\hfill
	\begin{subfigure}[t]{.49\linewidth}
		\includegraphics[width=\linewidth]{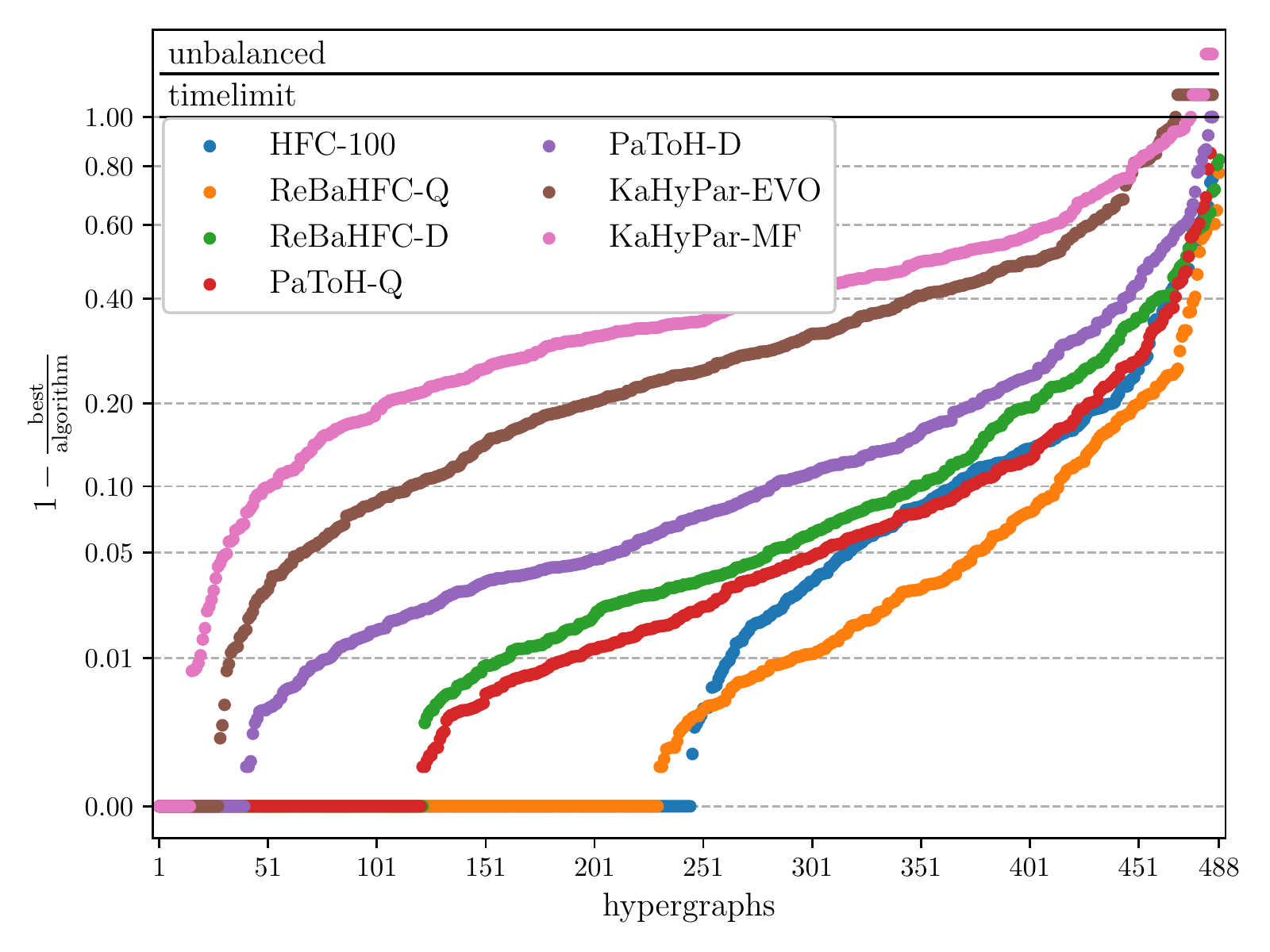}
	\end{subfigure}
	\caption{Comparison between the algorithms for $\epsilon = 0$. Left: Absolute running times for every hypergraph and random seed. Right: performance plot relating the minimum cut per algorithm and hypergraph to the overall best cut for that hypergraph. Lower values are better.}
	\label{fig:experimental:eps0.0}
\end{figure}

\subparagraph{Comparison for perfectly balanced partitioning.}
Even though the setting $\epsilon = 0$ has received no attention in hypergraph partitioning and only some attention in graph partitioning~\cite{ss-tlagh-13, mms-a-09, cbm-aprob-07, bh-aemma-10, bh-ammai-11, bh-aemts-11, dw-bbgb-12}, we consider it here.
Previous studies on perfectly balanced partitioning for graphs have focused on high quality solutions through running time intensive metaheuristics such as evolutionary algorithms \cite{ss-tlagh-13, bh-ammai-11, bh-aemma-10} or tabu search~\cite{bh-aemts-11} and even an exact branch-and-bound algorithm~\cite{dw-bbgb-12}.
Therefore, we include KaHyPar-EVO~\cite{ass-mmhp-18} (the evolutionary algorithm of KaHyPar) as well as plain HyperFlowCutter in addition to the already considered algorithms.
We exclude hMETIS-R from this comparison since it rejects $\epsilon < 0.002$ for bipartitions.

We include plain HyperFlowCutter with up to 100 terminal pairs as described in Section~\ref{sec:parameterstudy} and denote this configuration as HFC-100.
The evolutionary algorithm KaHyPar-EVO generates, manages and improves a pool of solutions until a time limit is exceeded, and outputs the minimum cut out of all generated solutions.
We set the instance-wise time limit to the maximum of the running times of HFC-100 and KaHyPar-MF to evaluate whether KaHyPar-EVO can yield better solution quality when given the same running time as HFC-100.
As opposed to the original paper, we configure KaHyPar-EVO to use flow-based refinement, which further improves solution quality.

KaHyPar-MF is unable to find any balanced bipartition on 4 hypergraps, whereas KaHyPar-EVO always finds one.
Furthermore, KaHyPar-MF exceeds the time limit on 7 hypergraphs and KaHyPar-EVO on an additional 17, without reporting intermediate solutions.
Figure~\ref{fig:experimental:eps0.0} shows the running times and a performance plot of all tested algorithms.
HFC-100 produces the best solutions on 245 hypergraphs, followed by ReBaHFC-Q (230), ReBaHFC-D (122), PaToH-Q (121), PaToH-D (40), KaHyPar-EVO (28) and finally KaHyPar-MF (15).
This shows that with exorbitant running time, HFC-100 produces high quality solutions for $\epsilon = 0$.
However the time-quality trade-off is clearly in favor of ReBaHFC-Q, especially since the solution quality of the latter is closer to the best cut for the instances on which it does not find the best cut, as opposed to HFC-100.
PaToH is better than KaHyPar for $\epsilon = 0$ because it includes a KL~\cite{kl-efppg-70} refinement pass as opposed to KaHyPar which only uses FM~\cite{fm-a-82}.\looseness=-1

\section{Conclusion}
In this paper we propose and evaluate HyperFlowCutter, a hypergraph bipartitioning algorithm based on maximum flow computations.
It enumerates partitions with increasing balance up to perfect balance.
We also propose and evaluate ReBaHFC, a refinement algorithm based on HyperFlowCutter.

In our experimental evaluation on a large set of hypergraphs, we show that while ReBaHFC is unable to beat the state-of-the-art hypergraph partitioners in terms of quality, it is still close in terms of quality and at the same time an order of magnitude faster.
Thus, it offers a new trade-off between quality and running time.
For the special case of perfectly balanced bipartitioning, the plain HyperFlowCutter algorithm, while being slow, computes the highest-quality solutions.
In this setting, ReBaHFC not only still beats all other partitioners but is also much faster.

In future work, it would be interesting to integrate the refinement step of ReBaHFC into multilevel partitioners to see if it can further improve their solution quality.


\bibliography{references}

\newpage

\appendix

\section{ReBaHFC Parameter Discussion}\label{sec:experimental:config_rebahfc}
In this section we discuss our parameter choices for ReBaHFC.
We use PaToH to obtain initial partitions for ReBaHFC, because it is between one and two orders of magnitude faster than KaHyPar and hMETIS, depending on whether the default or quality preset is used.

\begin{figure}
	\begin{subfigure}{.49\linewidth}
		\includegraphics[width=\linewidth]{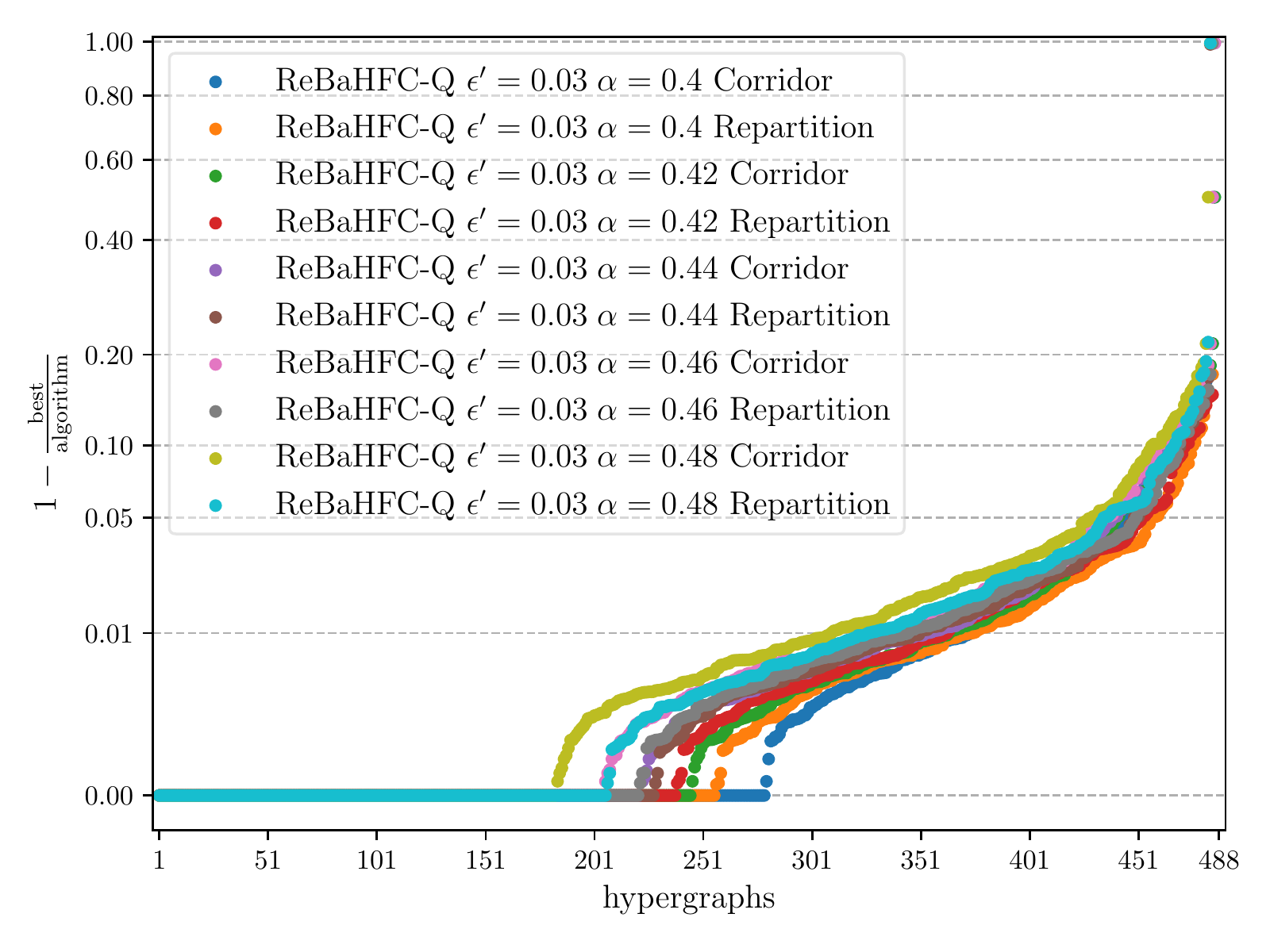}
		\caption{\centering ReBaHFC-Q, $\epsilon = 0.03$.}
	\end{subfigure}
	\hfill
	\begin{subfigure}{.49\linewidth}
		\includegraphics[width=\linewidth]{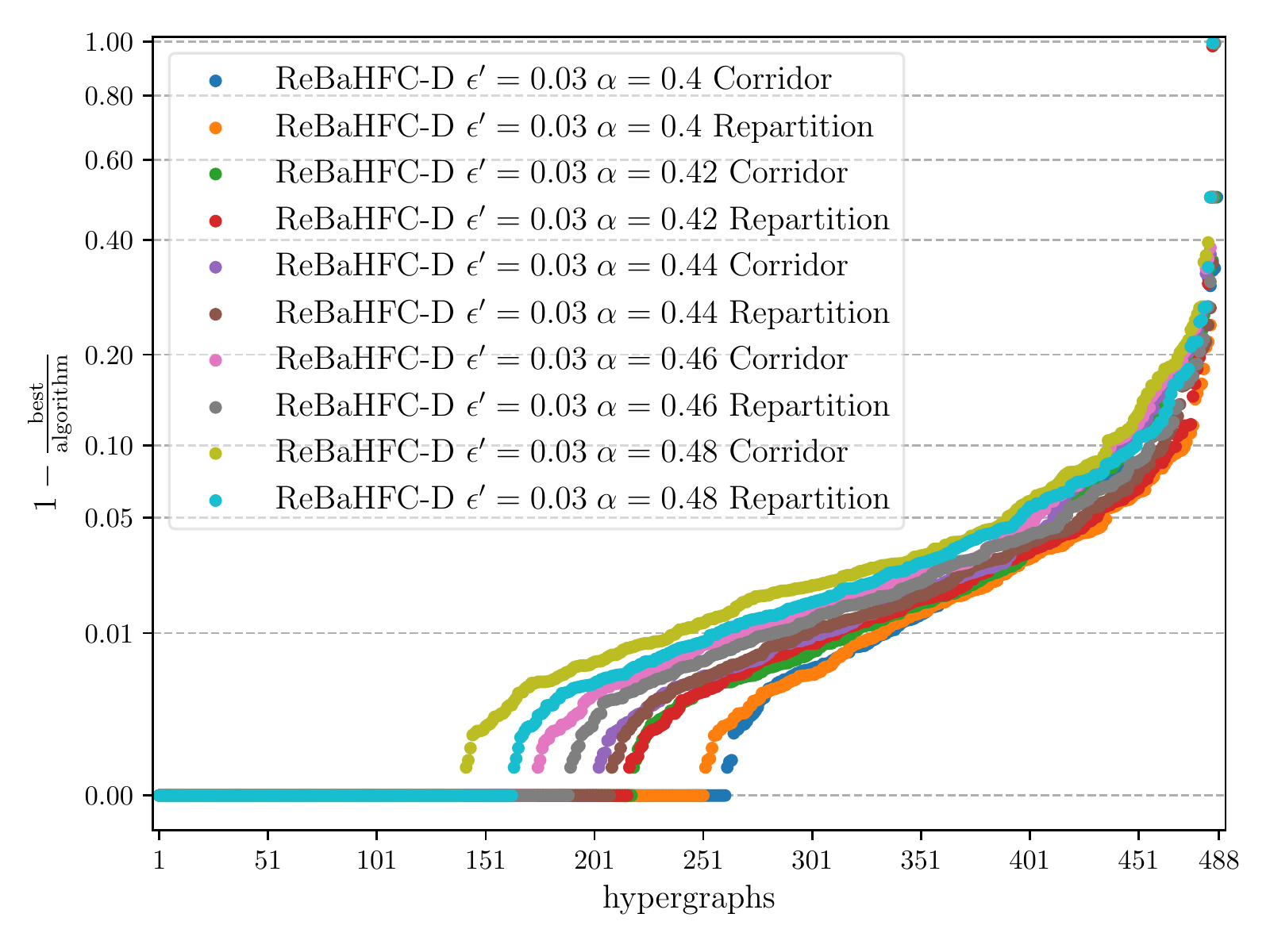}
		\caption{\centering ReBaHFC-D, $\epsilon = 0.03$.}
	\end{subfigure}
	\\
	\begin{subfigure}{.49\linewidth}
		\includegraphics[width=\linewidth]{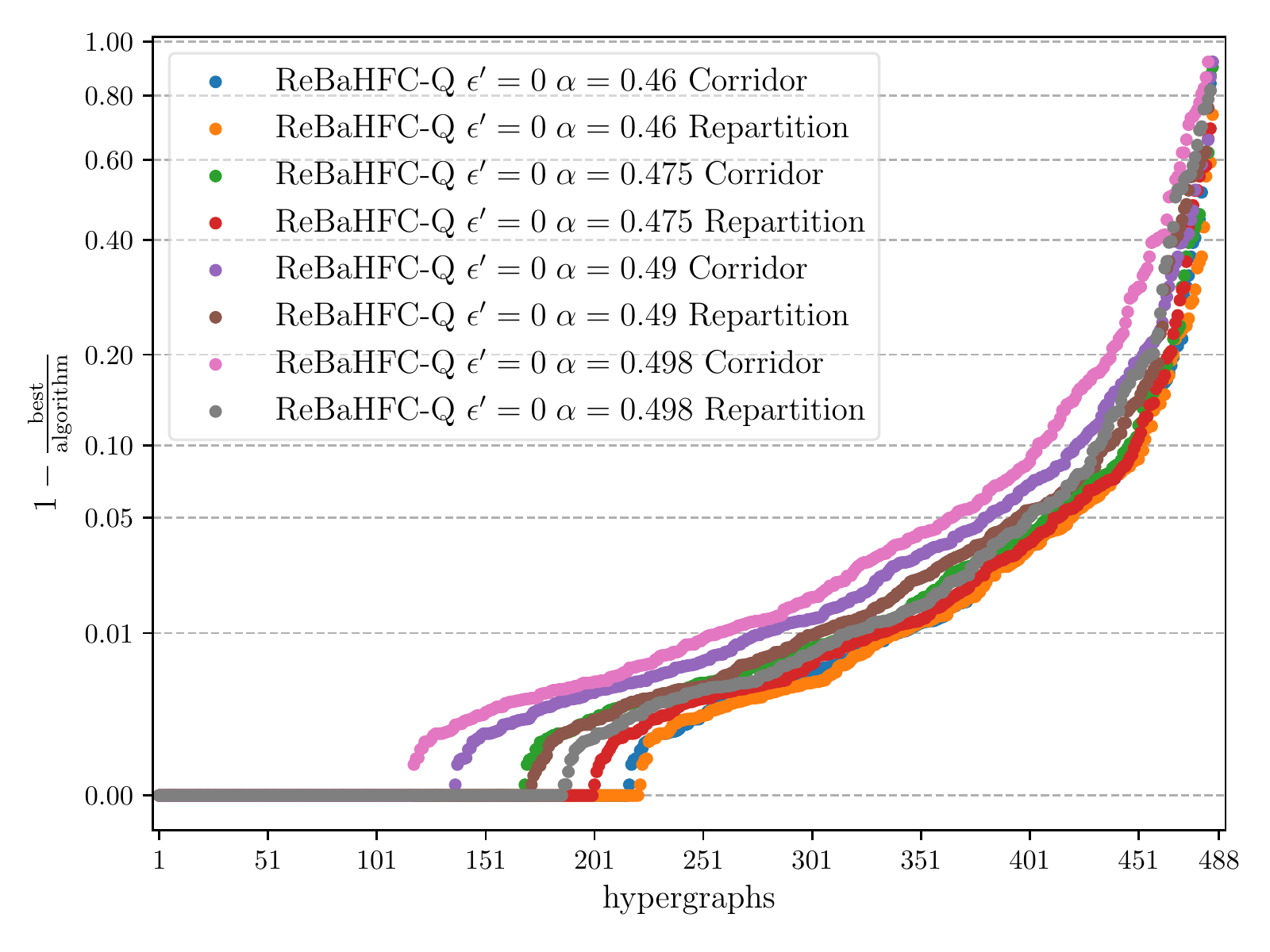}
		\caption{\centering ReBaHFC-Q, $\epsilon = 0$.}
	\end{subfigure}
	\hfill
	\begin{subfigure}{.49\linewidth}
		\includegraphics[width=\linewidth]{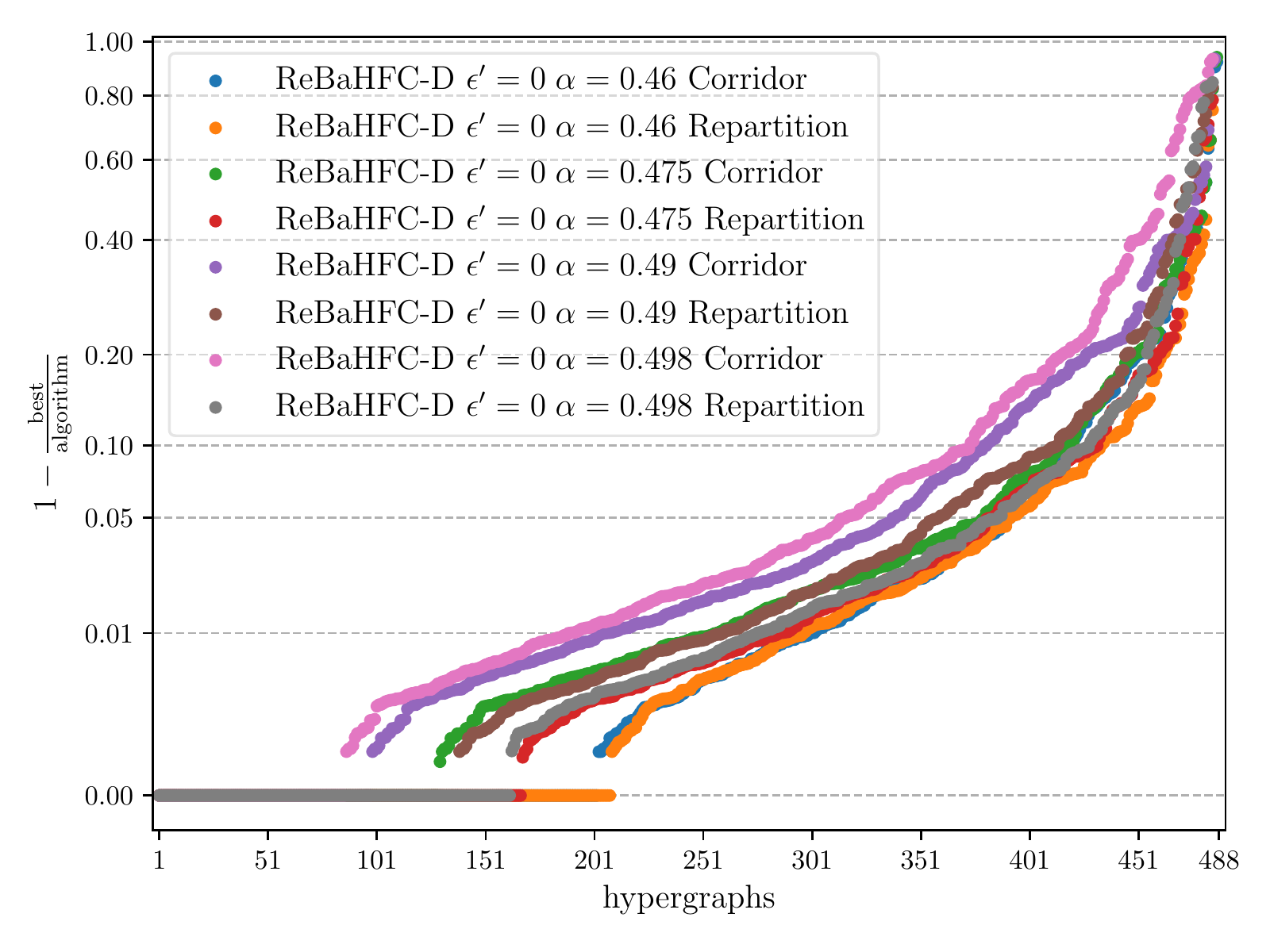}
		\caption{\centering ReBaHFC-D, $\epsilon = 0$.}
	\end{subfigure}
	\\
	\begin{subfigure}{.49\linewidth}
		\includegraphics[width=\linewidth]{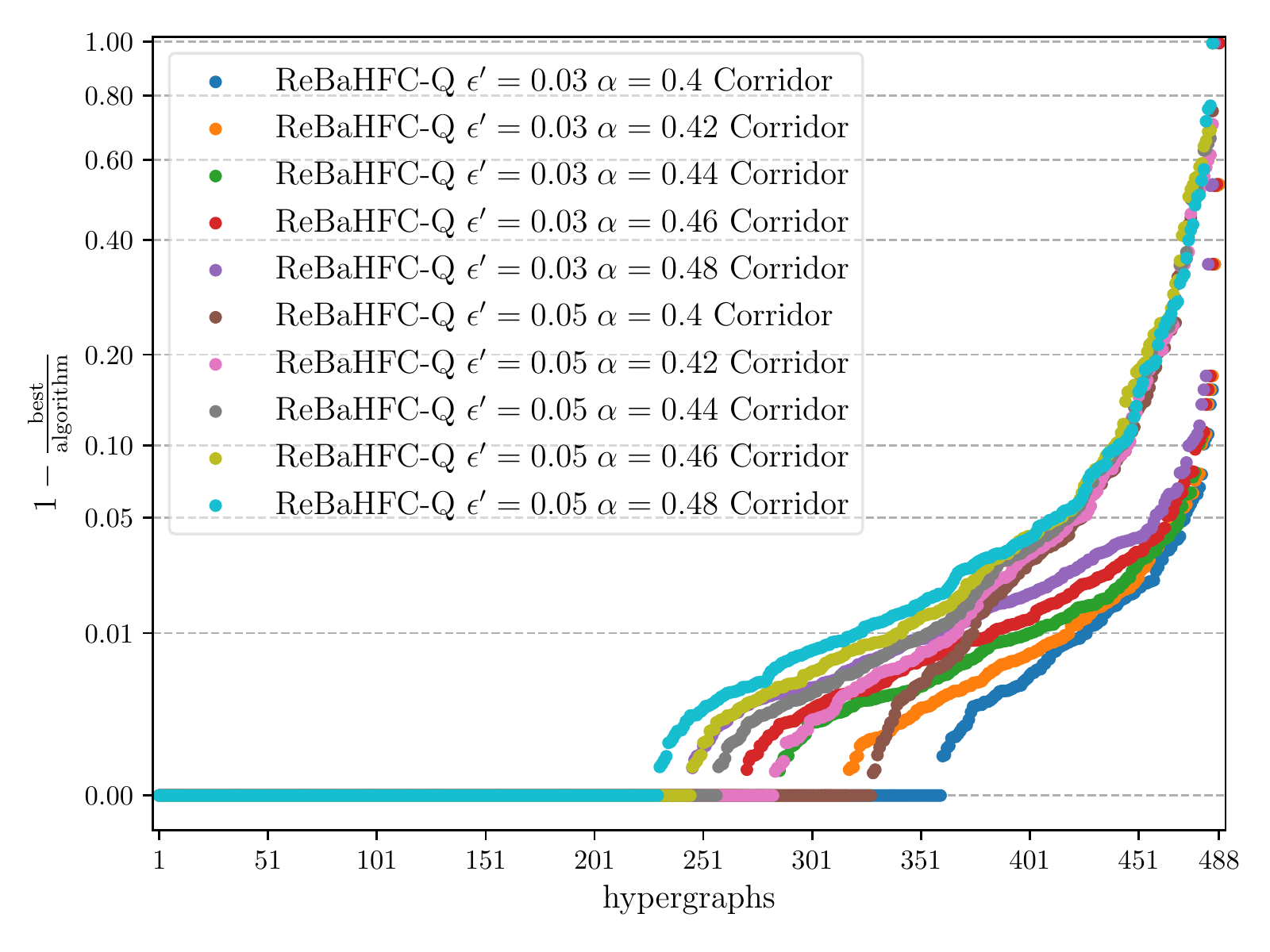}
		\caption{\centering ReBaHFC-Q with $\epsilon = 0.03$ and block resizing using corridors. This plot shows that it is better to use $\epsilon' = \epsilon$ for the initial partition.}\label{fig:larger_external_eps_does_not_help}
	\end{subfigure}
	
	\caption{Performance plots to compare the two block resizing strategies and block-size thresholds $\alpha$. For both PaToH presets we get the same best-performing block-size threshold $\alpha = 0.4$ for $\epsilon = 0.03$ and $\alpha=0.46$ for $\epsilon = 0$. The block resizing strategies barely differ in quality.}
	\label{fig:experimental:performanceplots_rebahfcconfiguration}
\end{figure}

In addition to constructing corridors using Breadth-First-Search, we also tried using PaToH to resize the blocks again.
Regarding solution quality, the two methods are roughly equivalent, which can be seen in Figure~\ref{fig:experimental:performanceplots_rebahfcconfiguration}.
However, we would like to investigate the suitability of HyperFlowCutter as a refinement algorithm in a multilevel framework such as KaHyPar, which is specifically suited since it already contains the necessary infrastructure for its own flow-based refinement.
Further, the necessary overhead of one or two additional invokations of PaToH led us to prefer corridor-construction using Breadth-First-Search.

We tested several block-size threshold values $\alpha$.
For $\epsilon = 0.03$ we tested $\alpha \in \{0.4, 0.42, 0.46, 0.48\}$ with $\alpha = 0.4$ working best.
For $\epsilon = 0$ we tested $\alpha \in \{0.46, 0.475, 0.49, 0.498\}$ with $\alpha = 0.46$ working best.
Smaller values for $\alpha$ are possible but are not recommended since larger flow problems would increase the running time too much.

Furthermore, we experimented with two different imbalance parameters $\epsilon' \in \{0.03, 0.05\}$ for the initial partition, with requested imbalance $\epsilon = 0.03$ for the output partition.
Figure~\ref{fig:larger_external_eps_does_not_help} shows that using an imbalance $\epsilon' > \epsilon$ for the initial partition yields worse output partitions.

\FloatBarrier

\section{Plain HyperFlowCutter Parameter Discussion}\label{sec:experimental:config_plain}
In this section we discuss the configuration choices for plain HyperFlowCutter.
Recall that we run plain HyperFlowCutter for $\epsilon = 0$ with up to 100 terminal pairs using interleaved execution in waves, as introduced in Section~\ref{sec:config:plain_hfc}.
All experiments in this section are for $\epsilon = 0$.
\subparagraph{Ensemble Terminals.}
\begin{table}
	\caption{Percentage of cases in the randomized experiment, which yield a better/worse partition, by replacing $Y$ randomly chosen, randomly generated terminal pairs out of 100 with the $Y$ first ensemble terminal pairs.}
	\label{table:ensemble}
	\centering
	\begin{tabular}{l  *5{r}}
		$Y$ & 1 & 3 & 5 & 7 & 10 \\ \midrule
		better [\%]  & +23.3 & +28.1 & +29.4 & +30.4 & +31.1 \\
		worse [\%] & -0.5 & -1.4 & -2.1 & -2.7 & -4.1 \\
	\end{tabular}
\end{table}

We evaluate the ensemble terminal pairs that were introduced in Section~\ref{sec:config:plain_hfc}.
To assess the impact of replacing randomly generated terminal pairs by ensemble terminal pairs, we propose a randomized experiment.
We replace $Y \in \{1,3,5,7,10\}$ randomly selected terminal pairs out of 100 by the $Y$ first ensemble terminal pairs and measure how often this improved the minimum cut.
The results in Table~\ref{table:ensemble} are accumulated over five runs of plain HyperFlowCutter with different random seeds, for each of which we accumulate 500 random samples of $Y$ ejected terminal pairs.
Any choice of $Y$ other than $1$ improves the solution in roughly 28-29\% (better $-$ worse) cases.
On 62.3\% of the hypergraphs, the first ensemble terminal pair is the best out of 10.
Thus we use only $Y=3$ ensemble terminal pairs for the final configuration.
This experiment was conducted on the 139 hypergraph parameter tuning subset used in Section~\ref{sec:parameterstudy}.

\begin{figure}[p]
	\centering
	\includegraphics[width=0.6\linewidth]{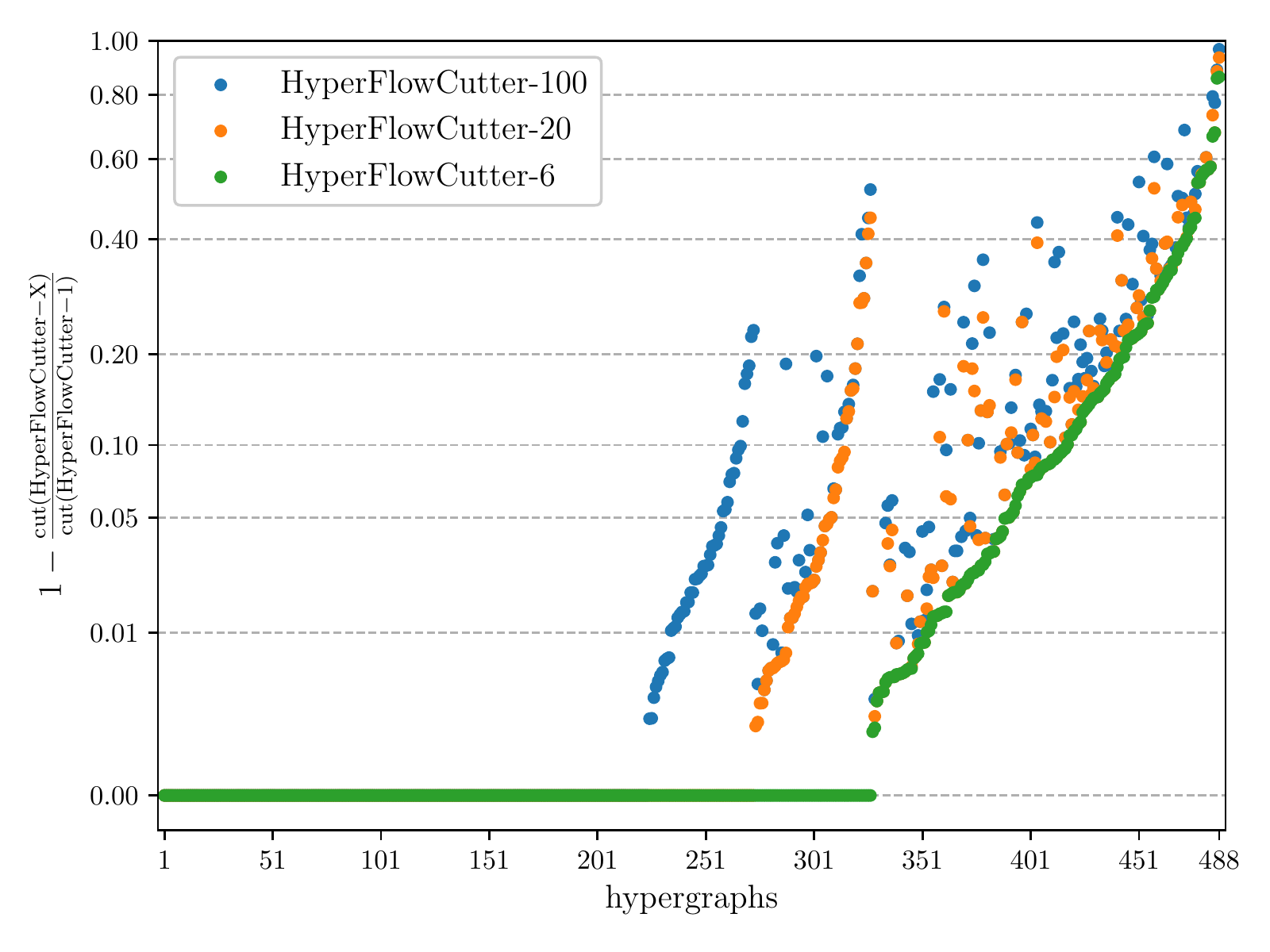}
	\caption{Ratios $1 - \frac{\cut{(\mathrm{HFC}-X})}{\cut{(\mathrm{HFC}-1)}}$ for the hypergraphs sorted lexicographically by the ratios of HFC-$6$, HFC-$20$ and then HFC-$100$. A ratio of zero means HFC-1 remains the best for this hypergraph, whereas higher values are better for HFC-$X$.}
	\label{fig:experimental:waves}
\end{figure}

\subparagraph{Waves and Impact of Number of Terminal Pairs.}
In addition to providing intermediate solutions in case the limit is exceeded, the interleaved execution in waves allows us to evaluate the impact of the number of terminal pairs on the entire benchmark set, without re-running the experiments.
We call the configuration, which takes the minimum of the first $\langle 1,2,3,4 \rangle$ waves finished within time limit, HFC-$\langle 1,6,20,100 \rangle$ respectively.
Figure~\ref{fig:experimental:waves} shows the sorted per-hypergraph ratios $1 - \frac{\cut{(\mathrm{HFC}-X)}}{\cut{(\mathrm{HFC}-1)}}$, which indicate by how much HFC-$\{6,20,100\}$ improved the bipartition of HFC-1.
A ratio of zero means HFC-X computed the same solution as HFC-1.
In this plot, the ratios share the same x-axis, since hypergraphs are sorted lexicographically by the ratios of HFC-$6$, HFC-$20$ and then HFC-$100$, as opposed to performance plots.
This is possible since adding terminals only improves the solution.
The purely green leg of 224 hypergraphs are instances on which HFC-1 still produces the best solutions.
This confirms the impact of ensemble terminal pairs on quality.
The green and blue leg from 225 to 259 are the instances where HFC-100 beats HFC-1 but HFC-20 and HFC-6 do not.
The first all-color leg from 260 to 311 are the instances where HFC-100 and HFC-20 beat HFC-6 but HFC-6 does not and on the remaining 177 instances all three compute smaller cuts than HFC-1.
\FloatBarrier

\section{Performance Plots for different Hypergraph Classes}\label{sec:experimental:performanceplots_instanceclasses}
\begin{figure}[H]
	\begin{subfigure}{.49\linewidth}
		\includegraphics[width=\linewidth]{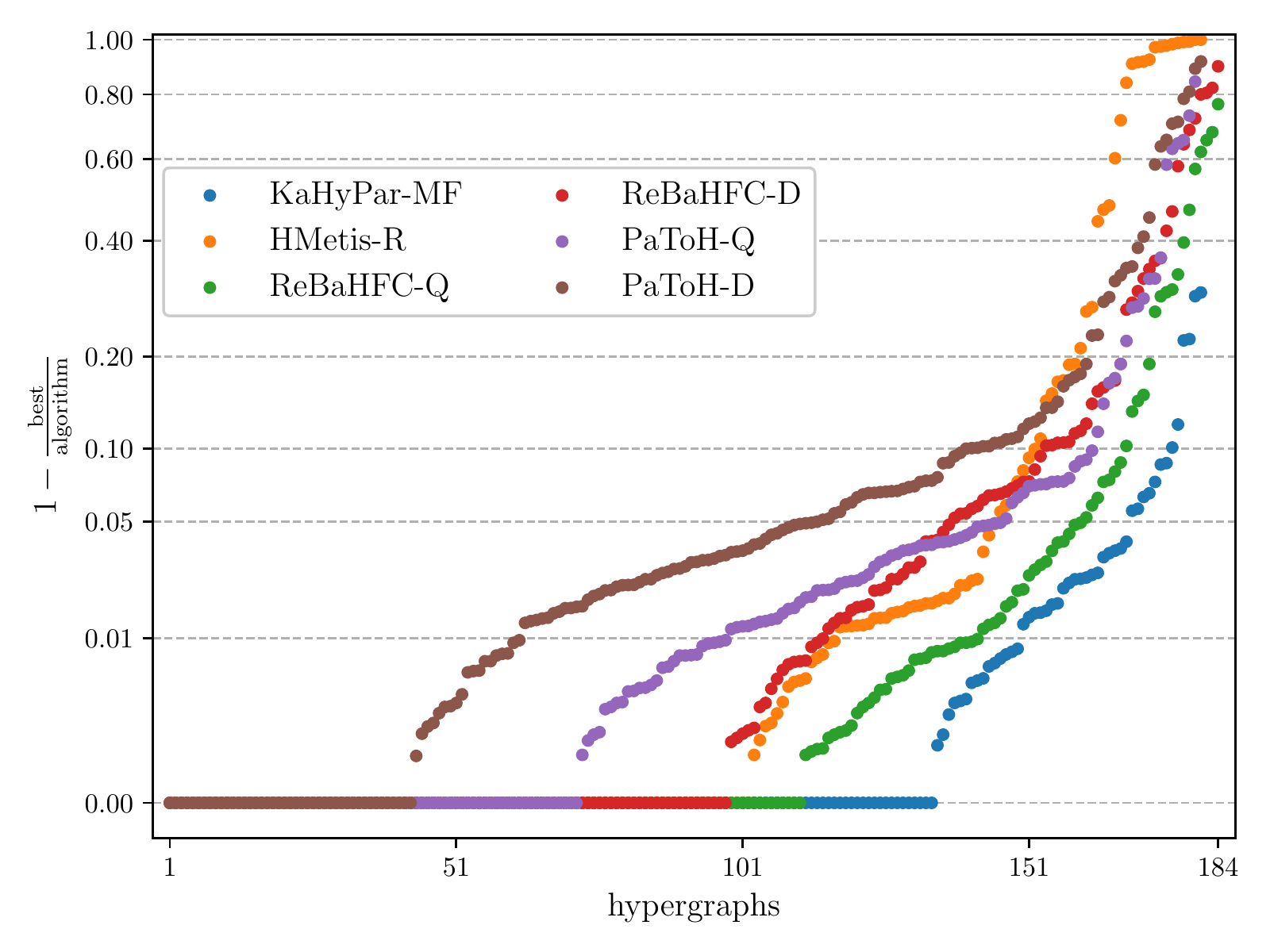}
		\caption{\centering Sparse Matrices.}
	\end{subfigure}
	\hfill
	\begin{subfigure}{.49\linewidth}
		\includegraphics[width=\linewidth]{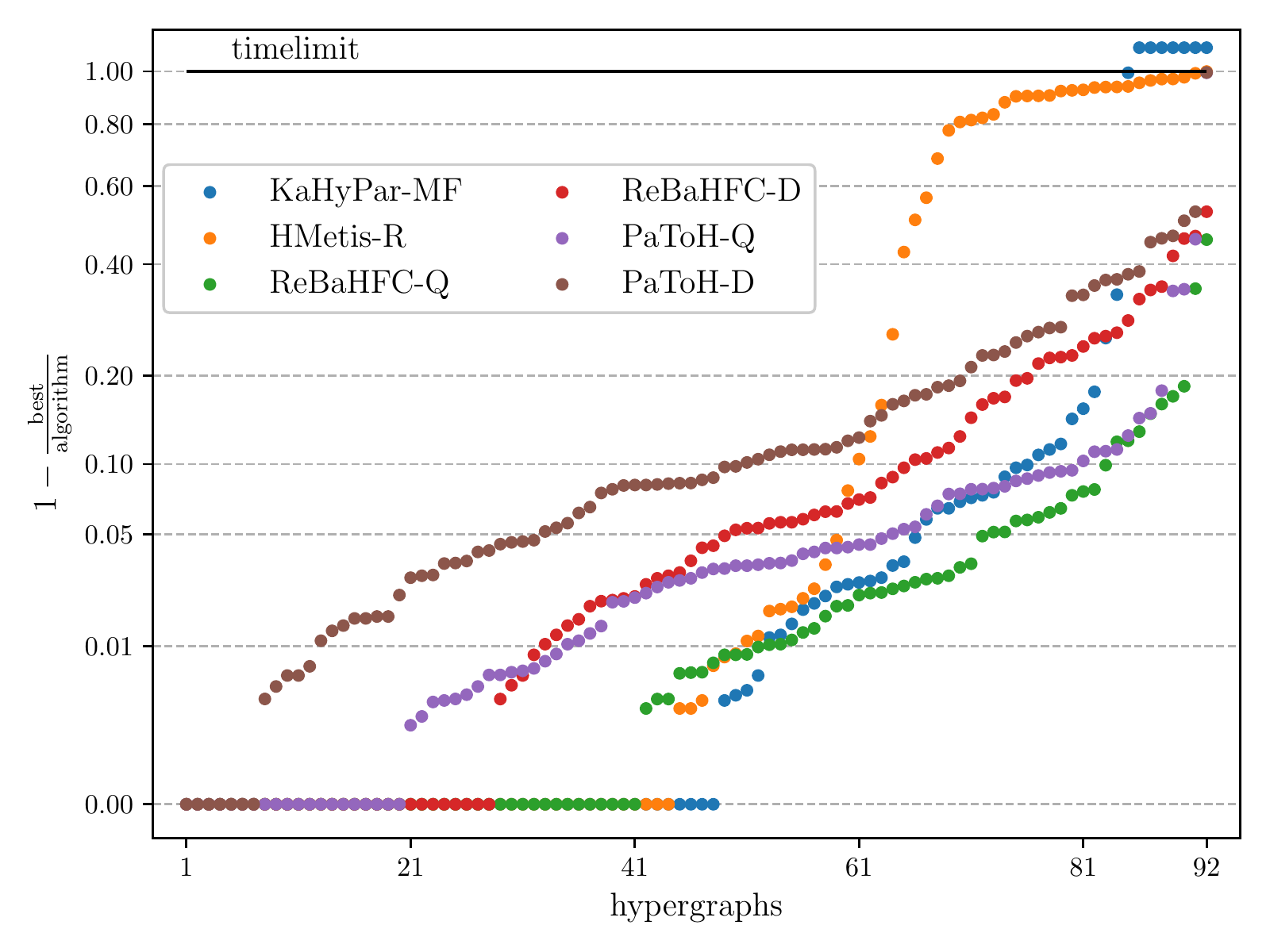}
		\caption{\centering Dual SAT.}
	\end{subfigure}
	\\
	\begin{subfigure}{.49\linewidth}
		\includegraphics[width=\linewidth]{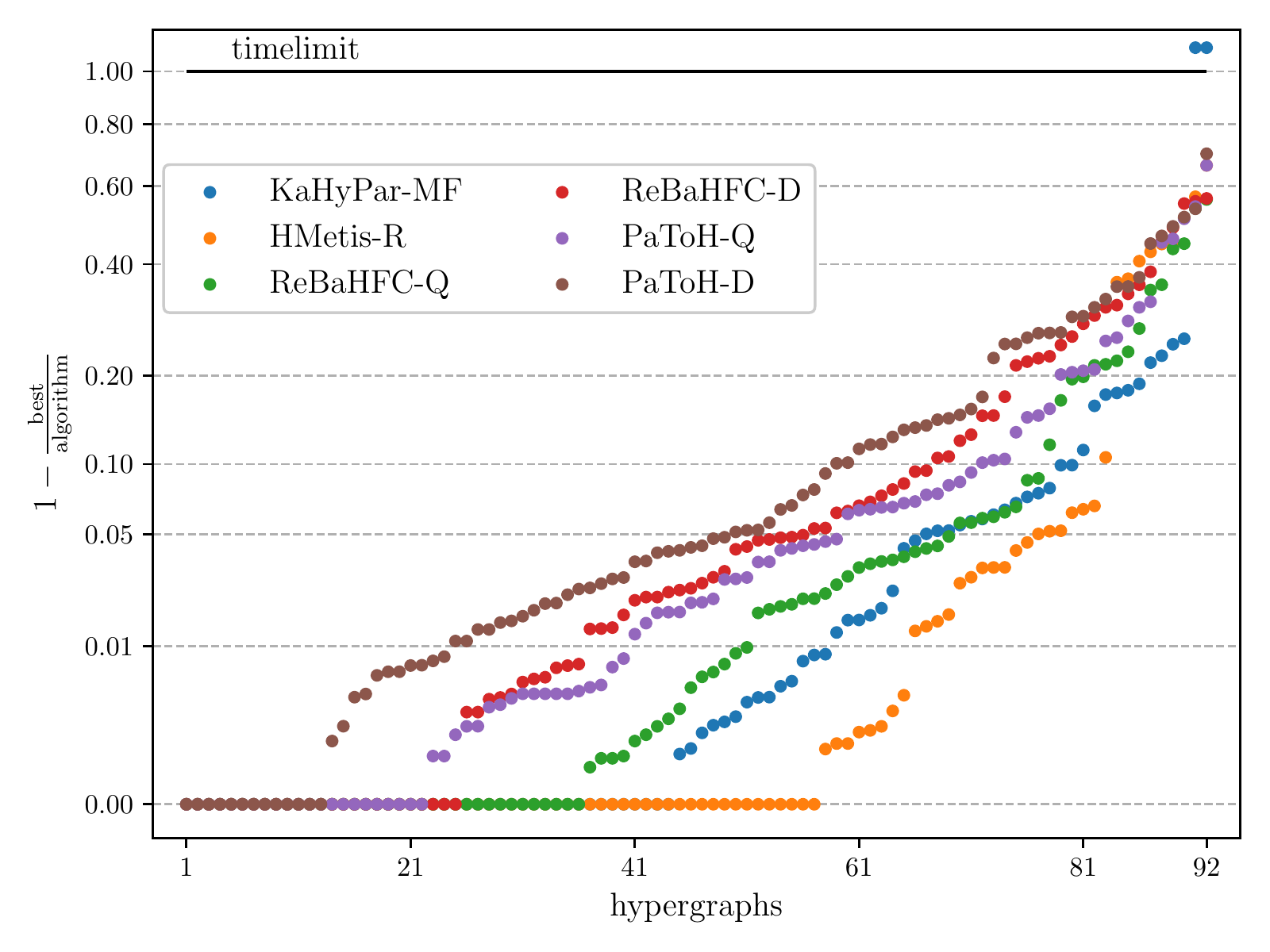}
		\caption{\centering Primal SAT.}
	\end{subfigure}
	\hfill
	\begin{subfigure}{.49\linewidth}
		\includegraphics[width=\linewidth]{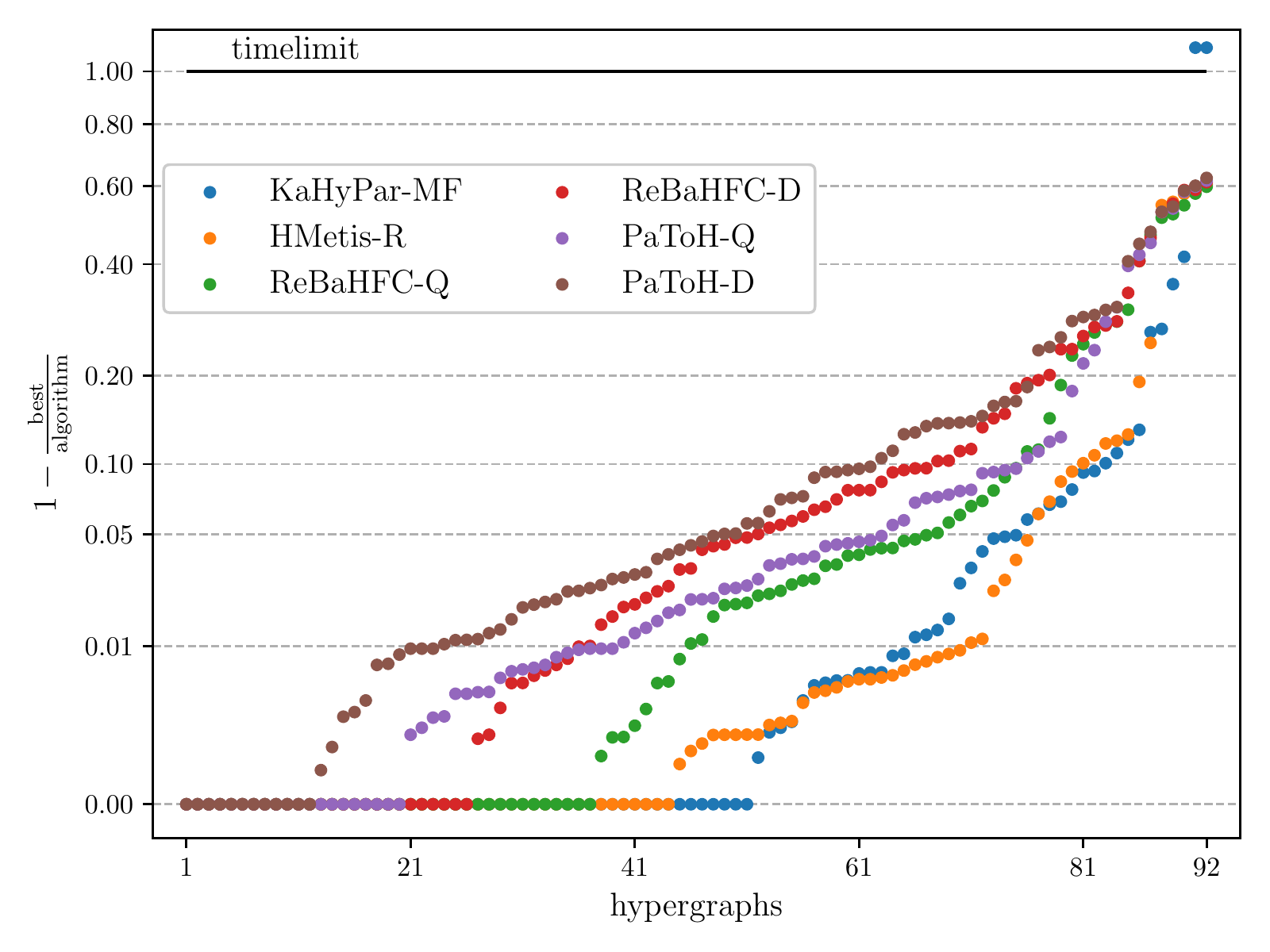}
		\caption{\centering Literal SAT.}
	\end{subfigure}
	\\
	\begin{subfigure}{.49\linewidth}
		\includegraphics[width=\linewidth]{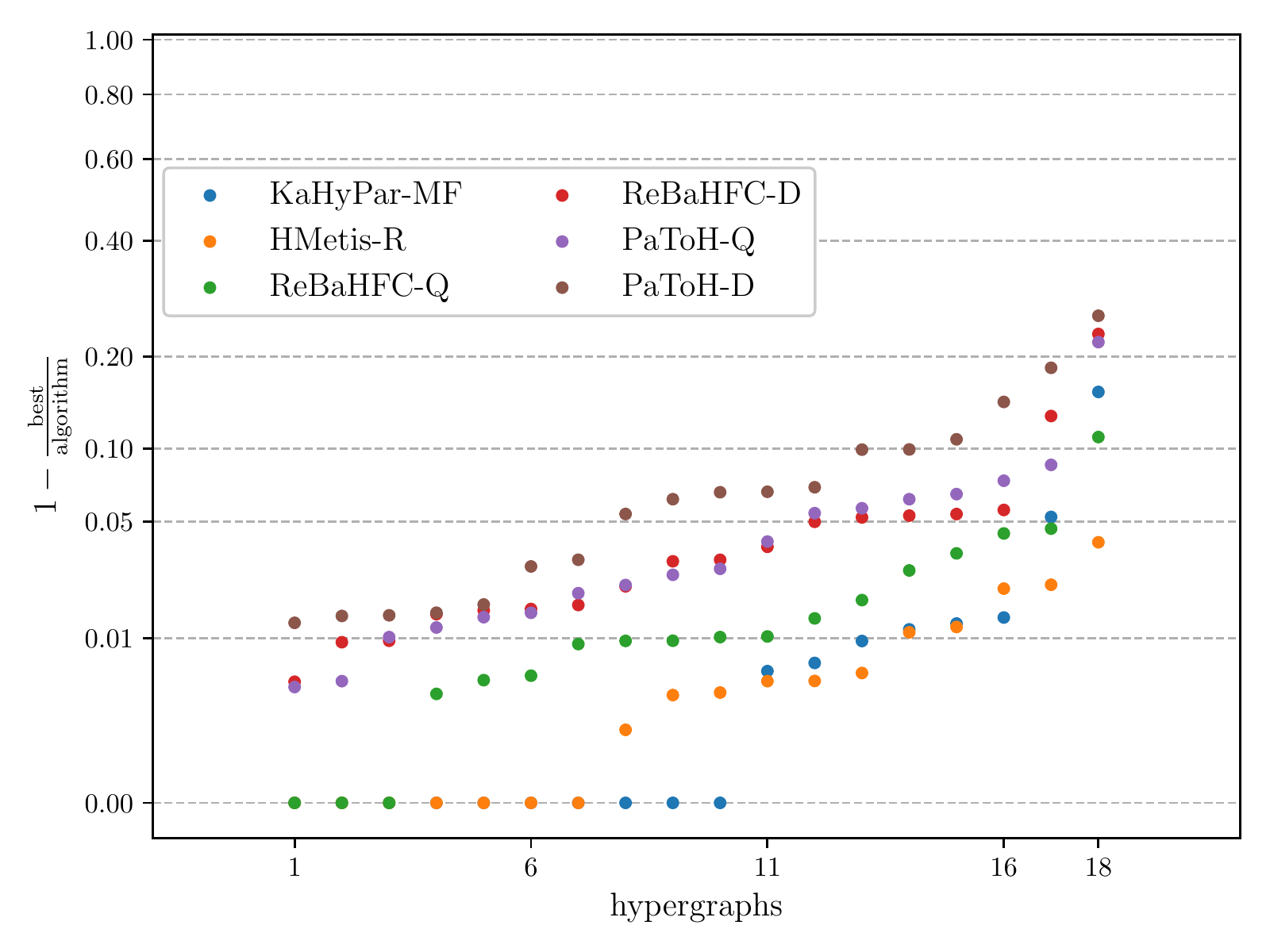}
		\caption{\centering ISPD98 VLSI.}
	\end{subfigure}
	\hfill
	\begin{subfigure}{.49\linewidth}
		\includegraphics[width=\linewidth]{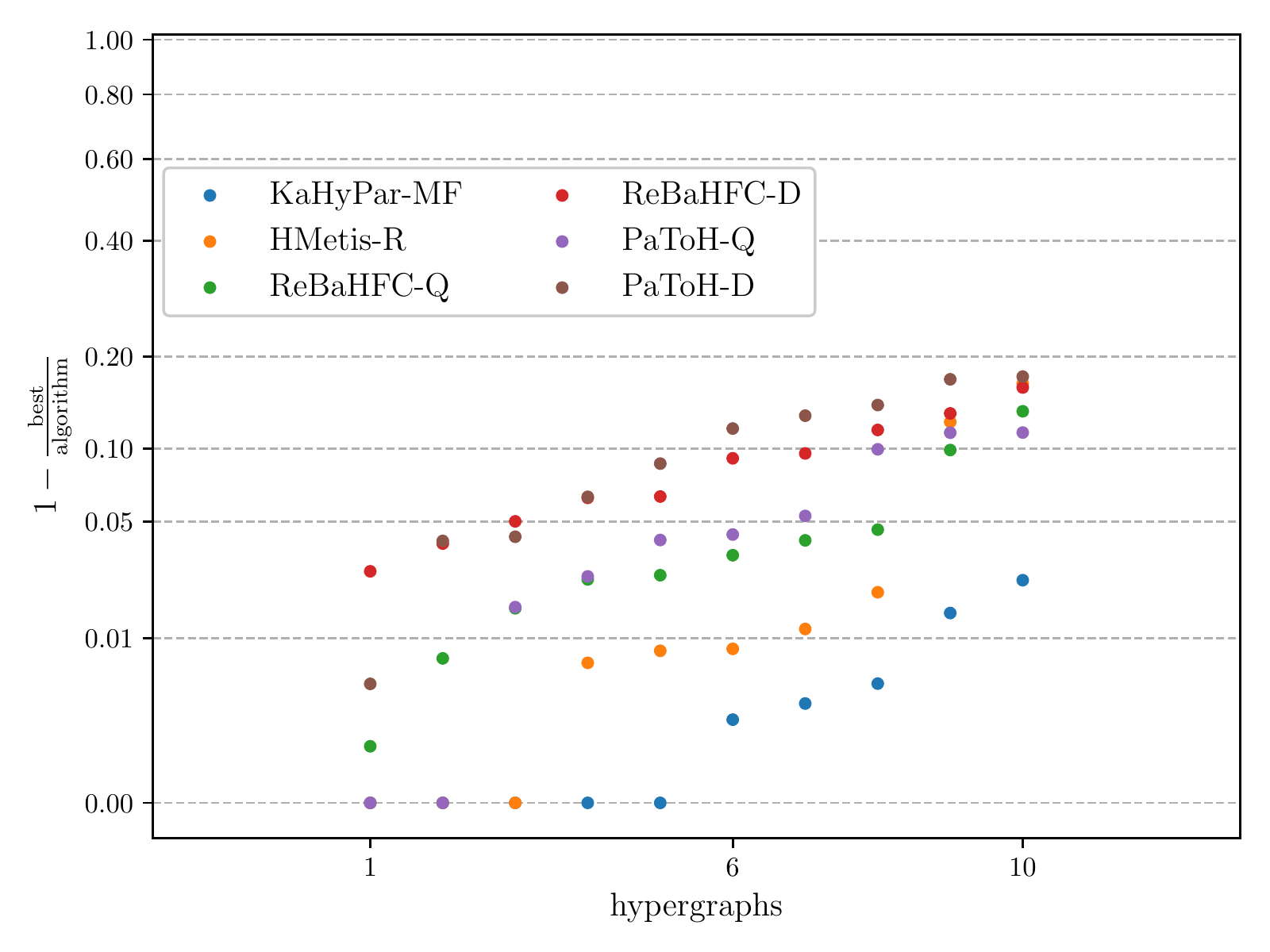}
		\caption{\centering DAC.}
	\end{subfigure}
	\caption{Performance plots for $\epsilon = 0.03$ comparing the algorithms on the different hypergraph classes of the benchmark set.}
	\label{fig:experimental:performanceplots_instanceclasses}
\end{figure}

\end{document}